\documentclass[useAMS]{mn2e}

\def\deg{^{\circ}}
\def\etal{{\it et al.}\thinspace}
\def\ie{{\it i.e.,}\thinspace}
\def\eg{{\it e.g.,}\thinspace}
\usepackage{graphicx}
\usepackage{rotating}
\usepackage{subfigure}
\usepackage{multirow}

\usepackage{amssymb}
\def\eq{\begin{equation}}
\def\en{\end{equation}}

\def\lesssim{\raisebox{-0.3ex}{\mbox{$\stackrel{<}{_\sim} \,$}}}
\def\gtrsim{\raisebox{-0.3ex}{\mbox{$\stackrel{>}{_\sim} \,$}}}

\def\etal{{\it et al.}\thinspace}
\def\ie{{\it i.e.}}

\def\eg{{\it e.g.}}
\def\apj{{\it Ap.J.}\thinspace}

\def\apjs{{\it Ap.J. Suppl.}\thinspace}
\def\aap{{\it A\&A}\thinspace}
\def\mnras{{\it MNRAS}\thinspace}
\def\P3hat{{\mathaccent 94 P}_3}

\def\nat{{\it Nature}\thinspace}

\newcommand{\simgt}%
        {\,\hbox{\lower0.6ex\hbox{$\sim$}\llap{\raise0.6ex\hbox{$>$}}}\,}
\newcommand{\simlt}%
        {\,\hbox{\lower0.6ex\hbox{$\sim$}\llap{\raise0.6ex\hbox{$<$}}}\,}
\usepackage{rotating}   

\title[Nulls, modes and interpulse of PSR B1944+17]{On the long and short nulls, modes and interpulse emission of radio pulsar B1944+17}

\author[Isabel M. Kloumann \& Joanna M. Rankin] 
{Isabel M. Kloumann$^{1}$ \& Joanna M. Rankin$^{2,1}$ \\ 
$^1$Physics Department, University of Vermont, Burlington, VT 05405\thanks{Isabel.Kloumann@uvm.edu; Joanna.Rankin@uvm.edu}  \\
$^2$Sterrenkundig Instituut `Anton Pannekoek', University of Amsterdam, NL-1098 XH\\
}

\date{Accepted 2009 month day. Received 2009 month day; in original form 2009 month day}

\pagerange{\pageref{firstpage}--\pageref{lastpage}} \pubyear{2009}

\def\LaTeX{L\kern-.36em\raise.3ex\hbox{a}\kern-.15em    T\kern-.1667em\lower.7ex\hbox{E}\kern-.125emX}

\begin{document}

\label{firstpage}

\maketitle

\begin{abstract}
We present a single pulse study of pulsar B1944+17, whose non-random nulls 
dominate nearly 70\% of its pulses and usually occur at mode boundaries. 
When not in the null state, this pulsar displays four bright modes of emission, 
three of which exhibit drifting subpulses.  B1944+17 displays a weak interpulse 
whose position relative to the main pulse ($\Delta\phi_{IP-MP}$) we find to be 
frequency independent. Its emission is nearly 100\% polarized, its polarization-angle 
traverse is very shallow and opposite in direction to that of the main pulse, 
and it nulls  approximately two-thirds of the time.  Geometric modeling indicates that 
this pulsar is a nearly aligned rotator whose $\alpha$ value is hardly 2\degr---\ie, its 
magnetic axis is so closely aligned with its rotation axis that its sightline orbit remains 
within its conal beam.  The star's nulls appear to be of two distinct types:  those with 
lengths less than about 8 rotation periods appear to be {\it pseudo}nulls---that is, 
produced by ``empty'' sightline traverses through the conal beam system; whereas 
the longer nulls appear to represent actual cessations of the pulsar's emission 
engine.  
\end{abstract} 


\maketitle

\section{Introduction}
\label{sec:intro}

Pulsar B1944+17 was discovered in August 1969 at the Molonglo Radio 
Observatory.  This 440-ms pulsar attracted attention thereafter because 
of its long null intervals, (Backer, 1970). Remarkably, it nulls some 70\% of the time and 
exhibits null lengths ranging between 1 and 300 stellar-rotation 
periods (hereafter $P_1$).  In 1986 Deich \etal\ (hereafter DCHR) investigated 
B1944+17's nulling behavior based on the received notion that its nulls 
represented ``turn offs'' of the pulsar emission mechanism. Thus they were 
concerned with the time scales of the cessations and resumptions of emission.  
While some pulsars, indeed, do appear to ``turn off'' for extended intervals---notably 
B1931+24 (Kramer \etal\ 2006)---there is a growing body of evidence that many 
nulls do not represent a shutdown of a pulsar's emission engine.  

Specifically, we now know that conal beams are comprised of rotating ``carousels'' 
of subbeams, and in certain situations nulls (or what might better be called 
{\it pseudo}nulls) represent ``empty'' sightline passes through a subbeam carousel.  
Such carousel ``action'' was first demonstrated in pulsar B0943+10 (\eg, Deshpande 
\& Rankin 1999, 2001; Suleymanova \& Rankin 2009), and then {\it pseudo}nulling 
was identified in pulsars B2303+30 (Redman \etal\ 2005; B0834+06 (Rankin \& 
Wright 2007) and J1819+1305 (Rankin \& Wright 2008).  Furthermore, periodicites 
clearly associated with nulling were discovered in pulsar B1133+16 by Herfindal 
\& Rankin (2007) and then subsequently in a number of other stars (Herfindal \& 
Rankin 2009; hereafter HR07/09), and such periodicities are almost certainly 
carousel related.  

There remain a small number of pulsars, however---B1944+17 prominent among 
them---whose observed nulling effects are not easily ascribed to either mechanism.    
That is, neither emission cessations, marked by partial nulls with near instantaneous decay times,  nor ``empty'' sightline traverses through a rotating subbeam carousel, marked 
by null periodicites, provide any clear explanation.

Pulsar B1944+17 exhibits a complex combination of behaviors including both very 
short and very long nulls.  In addition, DCHR identified what appeared to be several 
distinct emission modes, denoted A-D.  And furthermore, Hankins \& Fowler (1986; 
hereafter HF86) discovered that B1944+17 has a weak interpulse that nulls in 
synchrony with its main-pulse region (hereafter IP/MP).  Synchronous nulling indicates, 
remarkably, that whatever mechanism is responsible for MP nulling is also controlling 
the IP emission.

The presence of both MP and IP emission raises vexing questions about the overall 
emission geometry of the star.  Several analyses of pulsar geometry (\eg, Lyne \& 
Manchester 1988; Rankin 1993a,b, hereafter R93a,b) have mentioned B1944+17, but as 
for virtually all such pulsars,\footnote{Until very recently that is; see Weltevrede \& 
Wright (2009) and Keith \etal\ (2010).} no credible model of its overall MP-IP emission 
geometry has yet been articulated.  In short, neither an ``opposite pole'' nor ``single 
pole'' IP geometry provides an obvious solution, so among the various other issues, 
this basic question is still open for B1944+17.   

When not in the null state, MP pulse sequences (hereafter PSs) exhibit four 
modes of emission, three of which are well defined drift modes. Burst lengths 
are as large as 100 pulses, indicating a non-random null-burst distribution. The 
strict organization of the principle drift mode (Mode A, to be defined below) is in 
stark contrast with the disorganized non-drifting burst mode (Mode D) as well as 
the pulsar's preponderance of null pulses. \\\\

The rich variety of PS effects exhibited by this pulsar complicates its analysis, 
as well as any effort at modeling its many phenomena. While B1944+17 exhibits 
so many different identifiable behaviors (organized drifting, nearly ``chaotic'' subpulse 
behavior, bright emission, short nulls, long nulls, etc.), what makes the star 
so compelling is that in any time interval of reasonable length, ($\sim$2000 
$P_1$), one {\it will} see each of these behaviors. This consistency indicates that 
the processes that produce such variable emission patterns are in some way 
repeating themselves.

This paper reports a new synthetic study of pulsar B1944+17.  We have 
conducted long, high quality polarimetric observations using the upgraded 
Arecibo telescope at both meter and decimeter wavelengths, carefully analyzed 
the star's emission and nulls on a PS basis, and reconsidered the emission 
geometry of its MP and IP.  \S 2 describes the observations, \S 3 details aspects 
of our analyses, \S 4 builds a geometrical model, and \S 5 presents our thorough 
null analysis. \S 6 then provides a summary and discussion of our results.

\section{Observations}
\label{sec2:II}
The observations used in our analyses were made using the 305-m Arecibo Telescope 
in Puerto Rico. The 327-MHz (P band) and 1400-MHz (L band) polarized PSs were 
acquired using the upgraded instrument together with its Gregorian feed system and 
Wideband Arecibo Pulsar Processor (WAPP\footnote{http://www.naic.edu/\textasciitilde 
wapp}) on 2006 August 19 and 2008 March 15, consisting of 7038 and 5470 pulses, 
respectively, see Table \ref{observations}. The auto- and cross-correlations of the channel voltages were three-level 
sampled and produced by receivers connected to orthogonal linearly polarized feeds 
(but with a circular hybrid at P band).  Upon Fourier transforming, sufficient channels 
were synthesized across a 25-MHz (100-MHz at L band) bandpass, providing resolutions 
of approximately 1 milliperiod of longitude. The Stokes parameters have been corrected 
for dispersion, interstellar Faraday rotation, and various instrumental polarization effects. 
At L band, four 100-MHz channels were observed with centers at 1170, 1420, 1520, and 
1620 MHz. Both of the observations encountered virtually no interference (hereafter RFI), 
except for the 1620 MHz channel at L band which was disregarded.  At L band, the lower 
three 100-MHz bands were appropriately delayed and added together to give a 300-MHz 
effective bandwidth.  The PPAs of the two observations are approximately absolute in that 
they have been corrected for both ionospheric and interstellar Faraday rotation.

\begin{table}
\begin{center}
\caption{Observations}
\begin{tabular}{ccccc}
\hline\hline
Band & MJD & BW     & Resolution & Pulses\\
         &           &(MHz) &     ($\deg$)    & (\#)\\
\hline
P band & 53966 & 25   & 0.31  & 7038\\
L band & 54540 & 300 & 0.21  & 5470\\
\hline
\label{observations}
\end{tabular}
\end{center}
\end{table}

\begin{figure*}
	\centering
	\subfigure[327-MHz PPA histogram comprised of 7038 pulses]
	{\includegraphics[width=80mm]{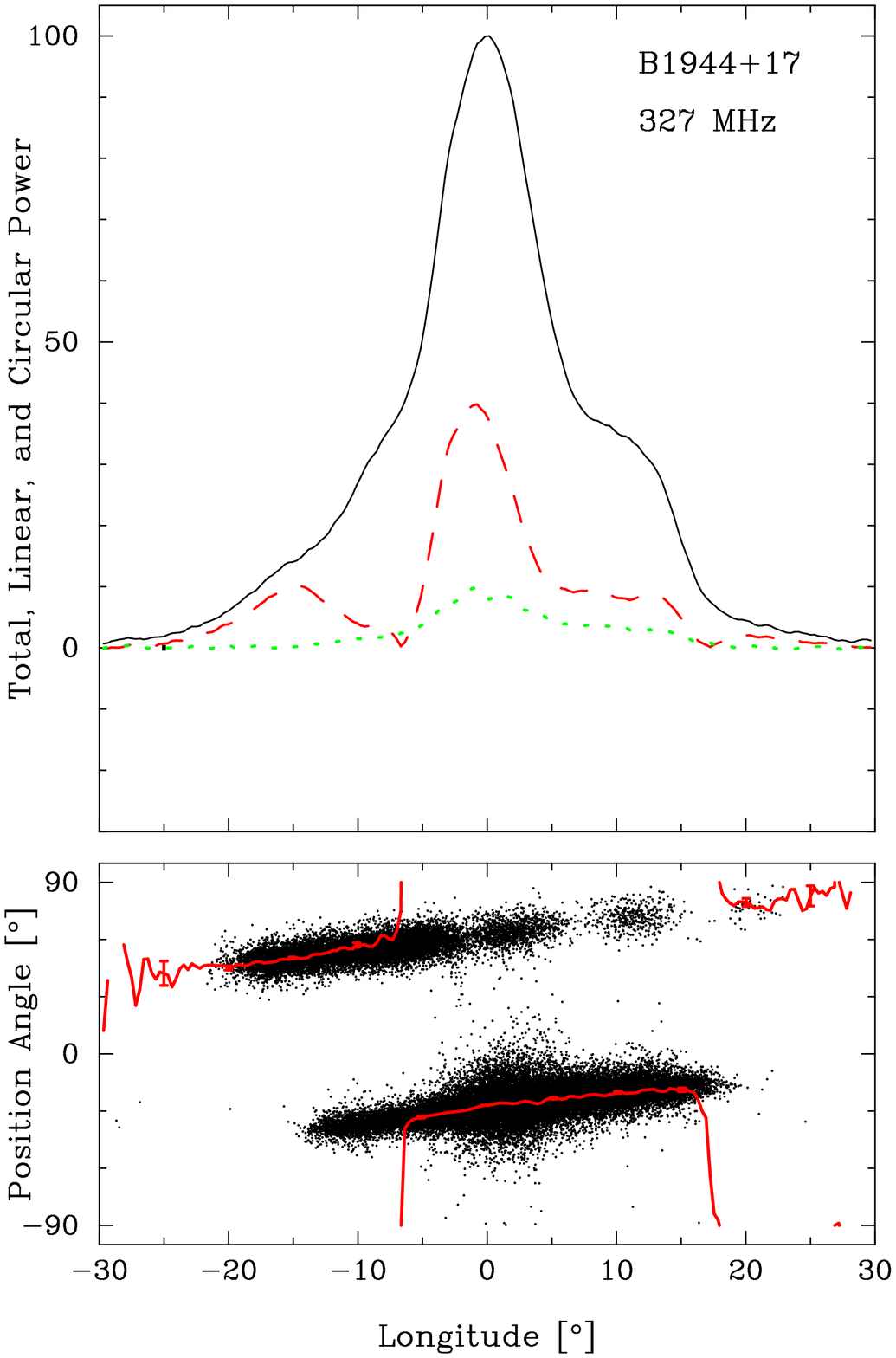}
	\label{327dottyplot}}
	\subfigure[1400-MHz PPA histogram comprised of 5470 pulses]
	{\includegraphics[width=80mm]{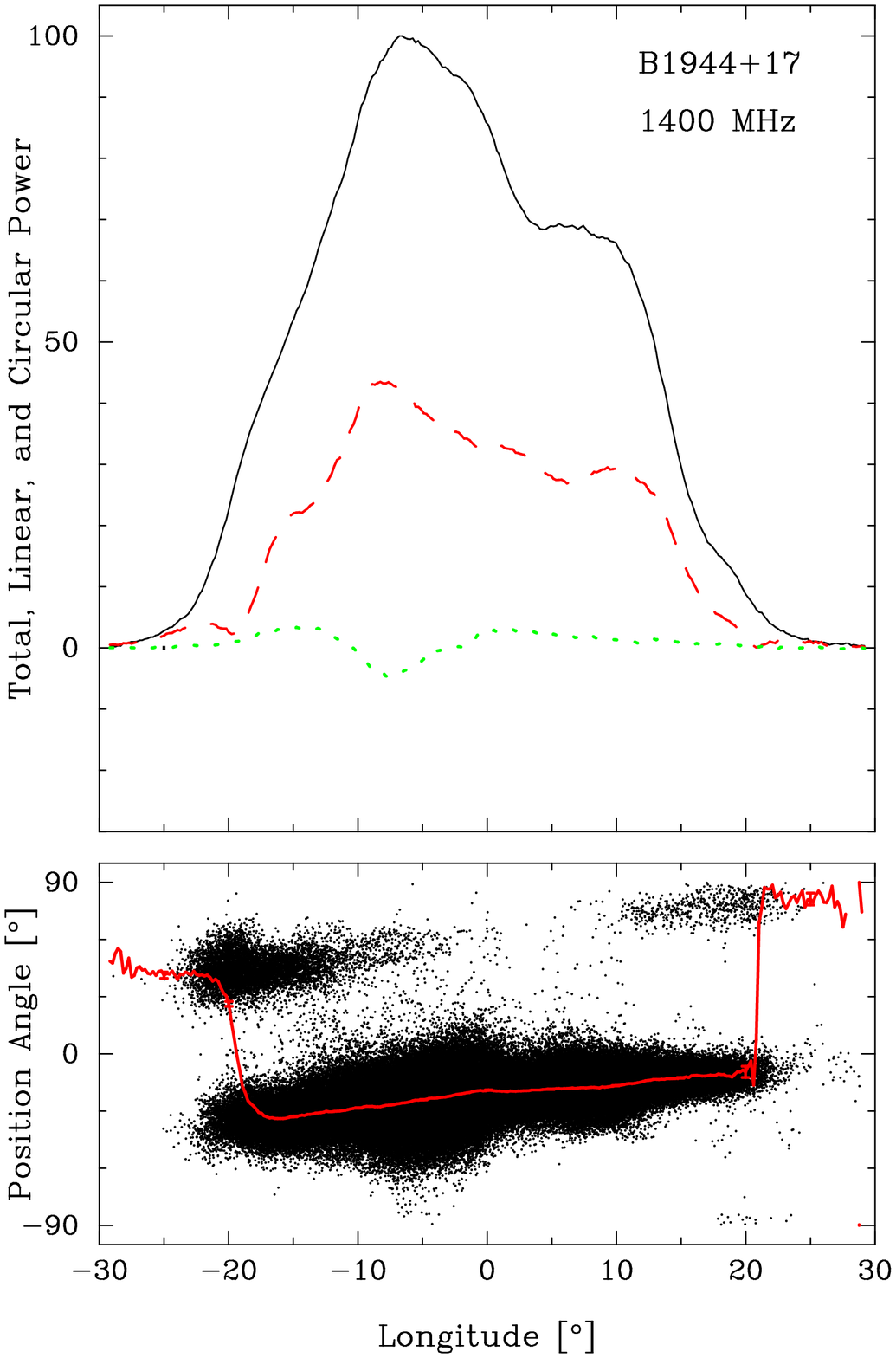}
	\label{1400dottyplt}}
\caption{The two panels display the total power (Stokes $I$), total linear 
polarization ($L$ [=$\sqrt{Q^2+U^2}$]; dashed red) and circular polarization 
(Stokes $V$, defined as left -- right-hand; dotted green) (upper), and the 
polarization angle ({\it PPA} [=$\frac{1}{2}\tan^{-1}(U/Q)$]) (lower).  Individual 
samples that exceed an appropriate $>$2 sigma threshold appear as dots 
with the average PPA (red curve) overplotted.  The tiny black box at the left 
of the upper panel gives the resolution and a deflection corresponding to 
three off-pulse noise standard deviations.  The PPAs are approximately 
absolute, such that the discontinuous regions of OPM emission at positive 
and negative PPAs correspond to each other.  Note that the P and L-band 
profiles each extend {\bf more than} $\pm$25\degr\ and that the lower 
frequency profile has an unusually narrow equivalent width relative to its 
higher frequency counterpart.}
\label{dottyplots}
\end{figure*}

\begin{figure}
\begin{center}
\includegraphics[width=66mm,angle=-90.]{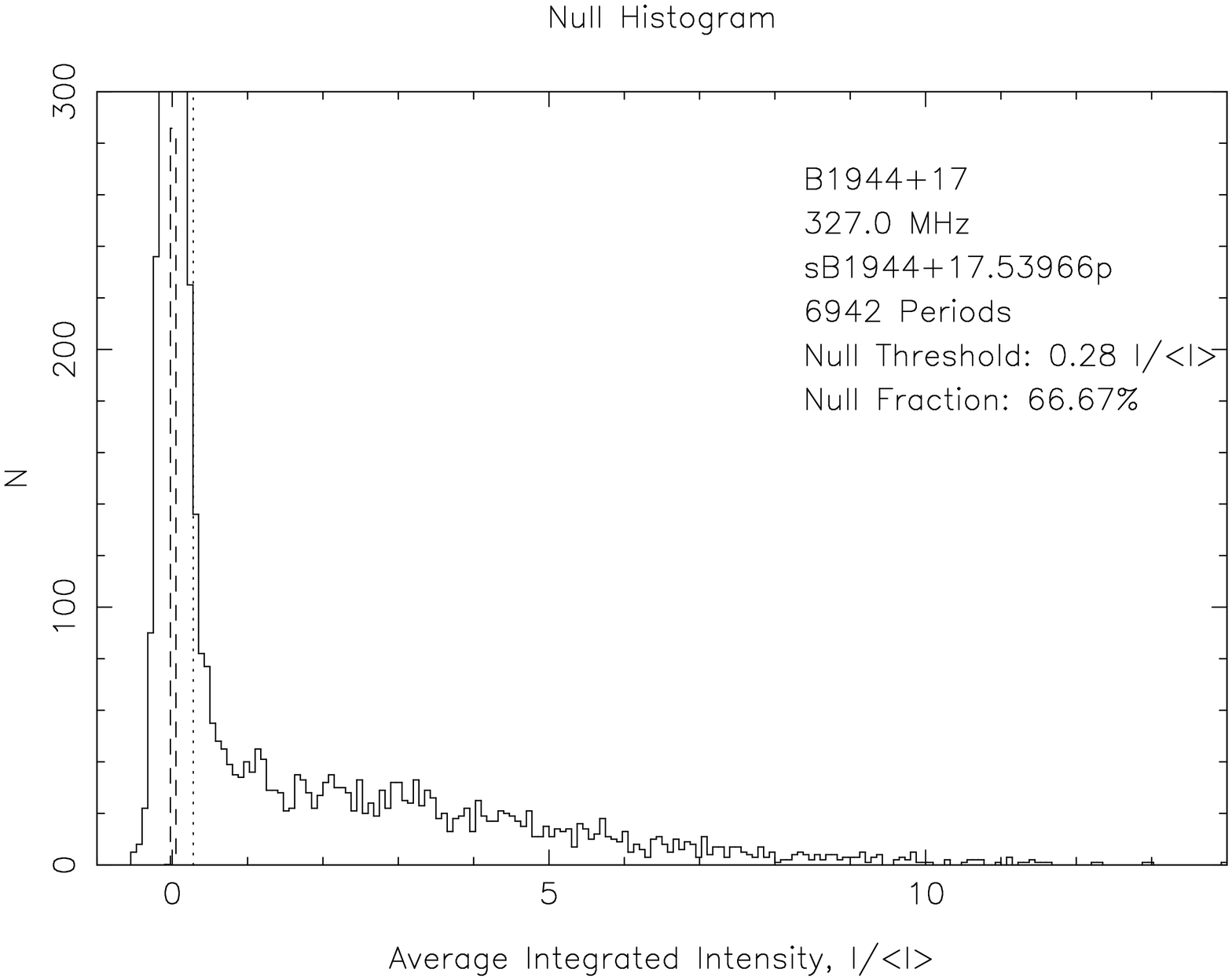}
\includegraphics[width=66mm,angle=-90.]{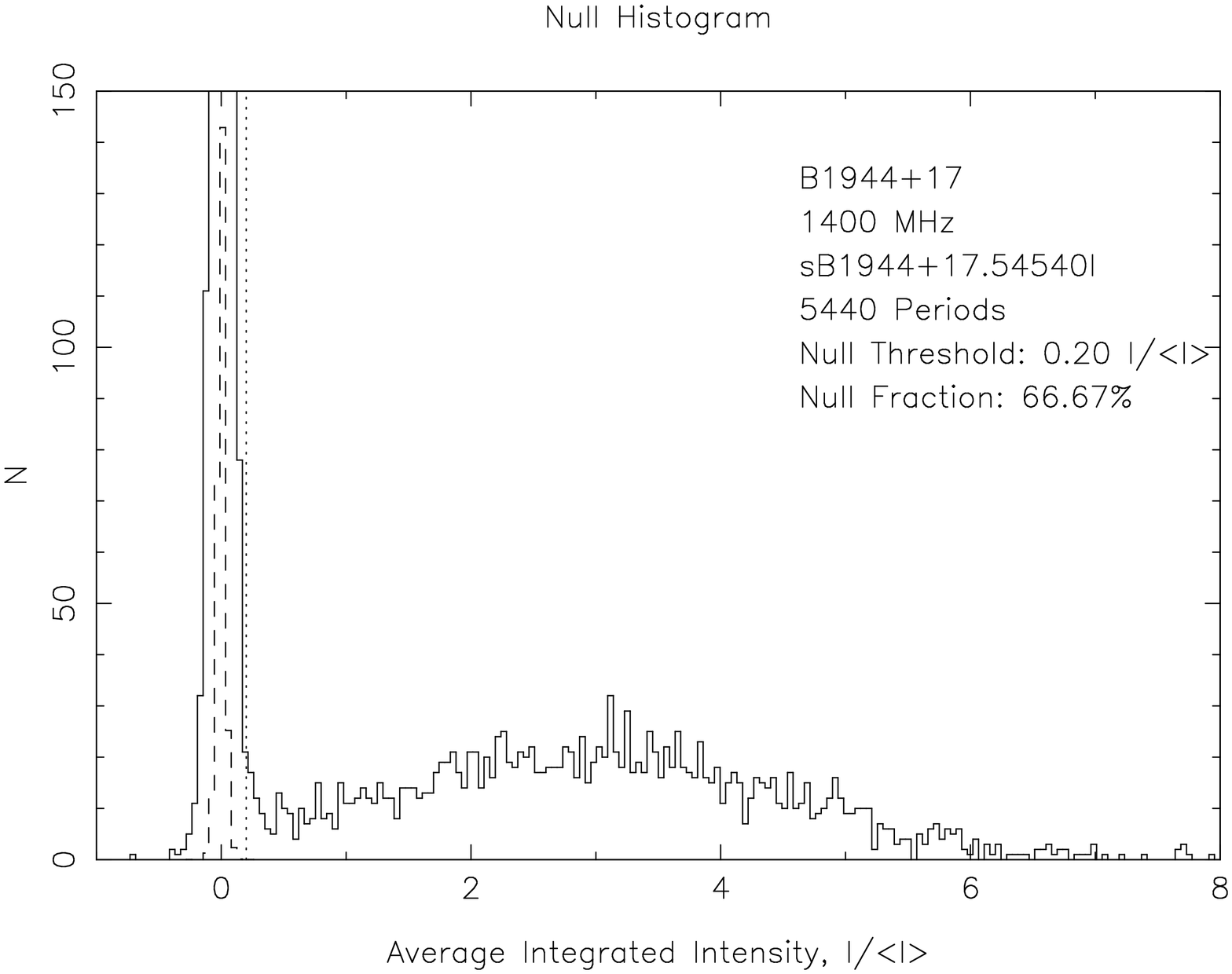}
\caption{Null histograms for B1944+17 at P (top) and L (bottom) bands.  
The histogram peaks (at 1110 and 980) corresponding to 
the large null fractions have been truncated in order to better show the 
pulse-amplitude distribution.  Despite the high S/N, the star's null- and 
pulse-energy distributions overlap, so that the nulls and pulses cannot  
be fully distinguished, though this difficulty is more severe at P band than 
at L band.  Plausible, conservative thresholds (shown by dotted vertical 
lines) of 0.28 and 0.20 $<$$I$$>$, respectively, indicate that some 2/3 
of the pulses are nulls or pseudonulls.}
\label{nullhistograms}
\end{center}
\end{figure}

\begin{figure*}
\begin{center}
\begin{tabular}{@{}lr@{}}
{\includegraphics[width=66mm,angle=-90.]{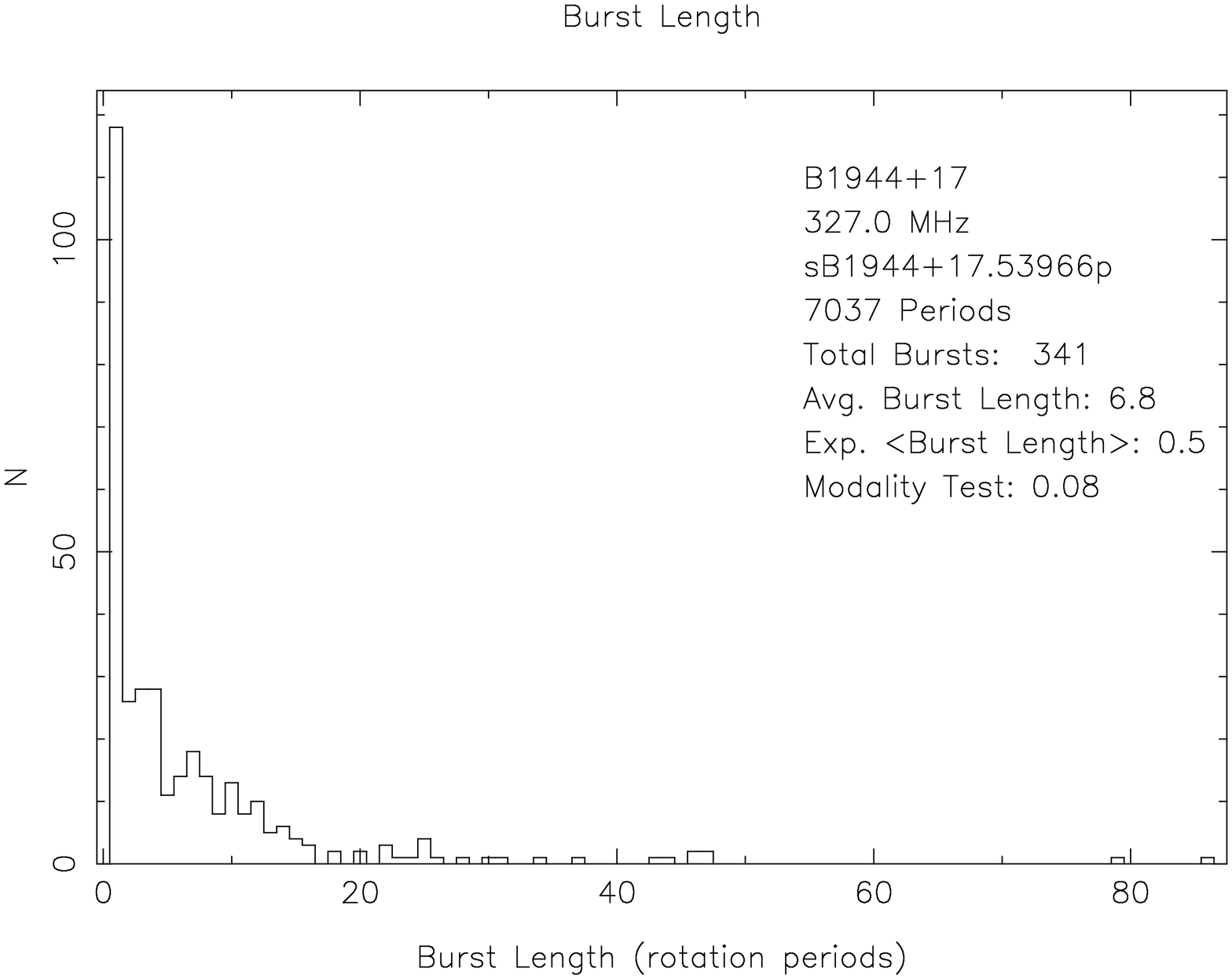}
\includegraphics[width=66mm,angle=-90.]{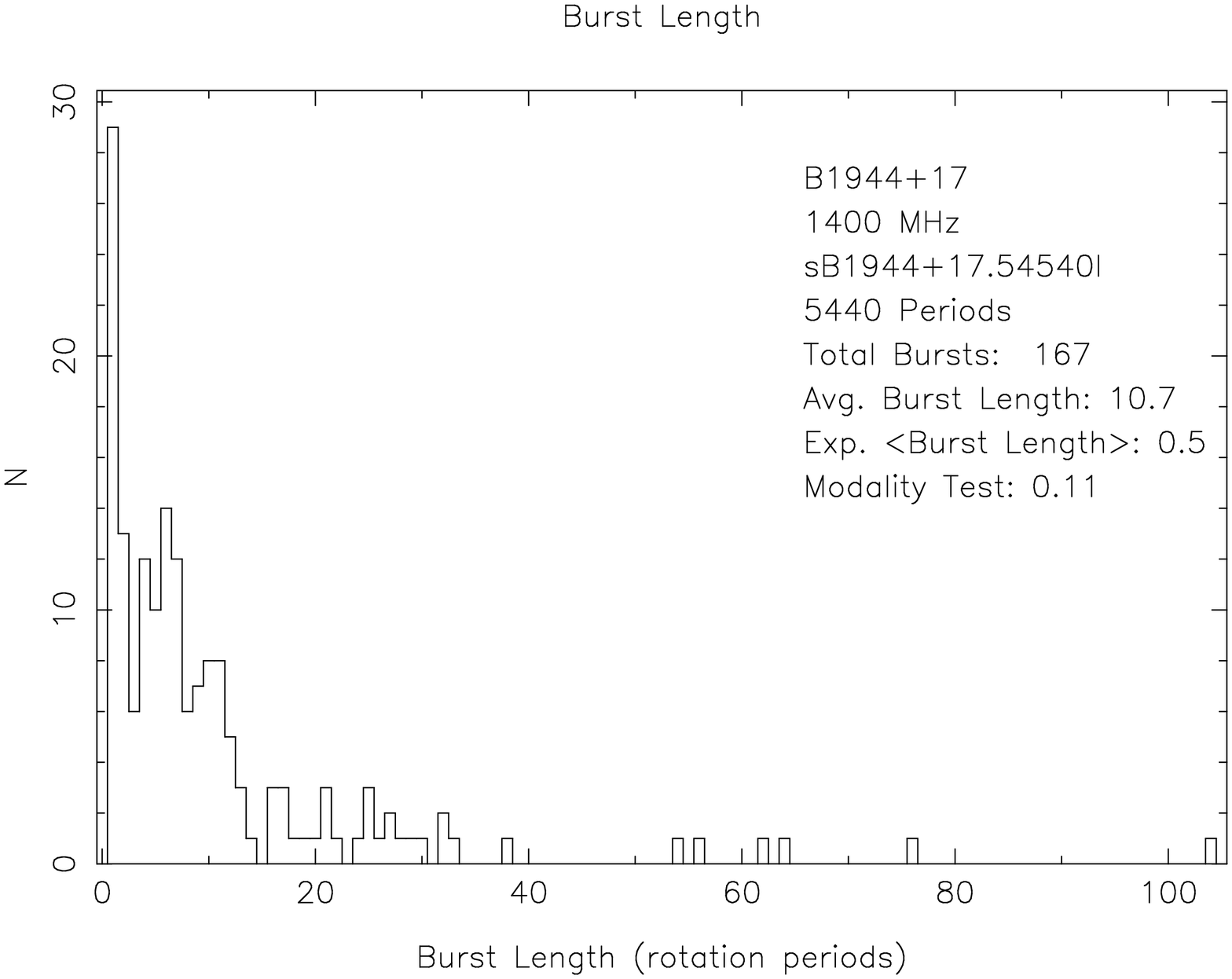}
	\label{Bursts}} \\
{\includegraphics[width=66mm,angle=-90.]{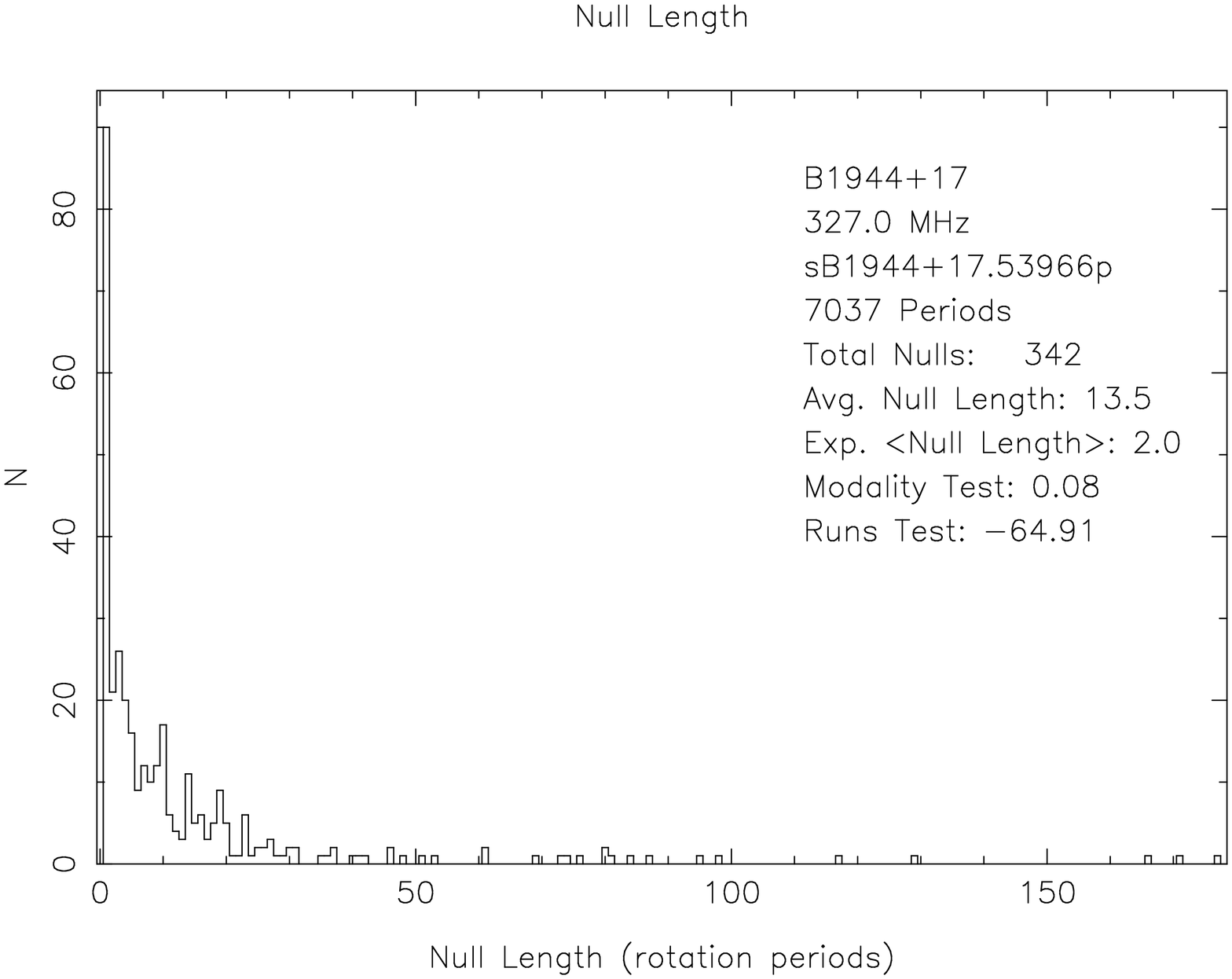}
\includegraphics[width=66mm,angle=-90.]{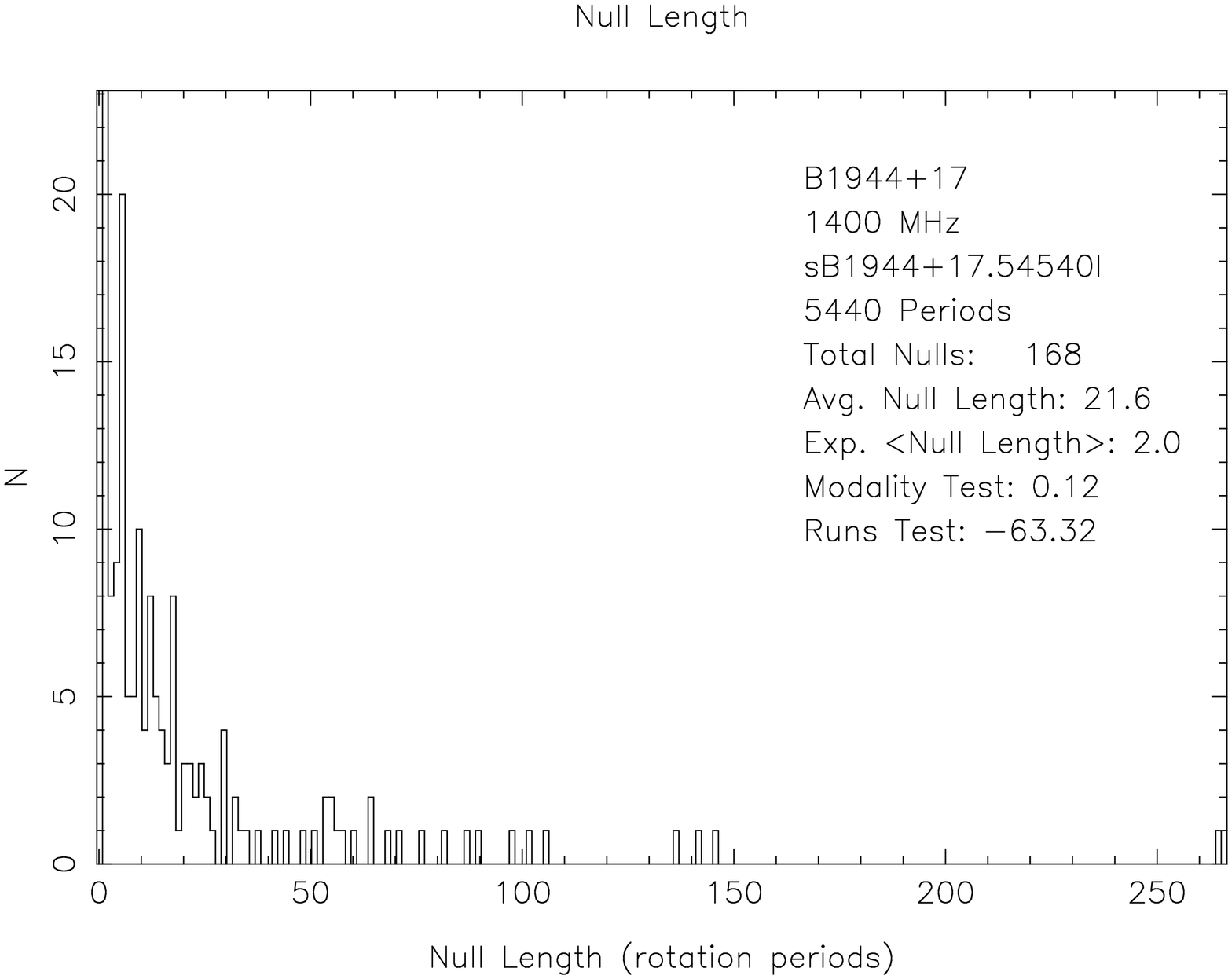}
	\label{Nulls}} \\
\end{tabular}
\caption{Burst- and null-length histograms corresponding to the P- and L-band 
observations and thresholds in Figure \ref{nullhistograms}.  These distributions 
show unsurprisingly that burst and null lengths of a single rotation period are 
highly favored; however, we also see that bursts can last up to approximately 90 
$P_1$ and nulls up to three times this! Thus, in the language of the Runs Test, 
the nulls occur non-randomly in the PS by virtue of being ``undermixed'' (see text). 
Note that the P-band observation included nulls up to 300 $P_1$; however the 
null histogram has been truncated down to 180, so as to resolve detail in the short 
null distribution.}
\label{burstnullfreq}
\end{center}
\end{figure*}

\section{Analyses}
Figure \ref{dottyplots} presents the polarized profiles and polarization-angle 
(hereafter PPA) histograms of pulsar B1944+17 at both 327 and 1400 MHz.  
While these profiles are familiar, it is useful to inspect them in detail.  Notice 
that the half-power or equivalent width of the star's roughly symmetrical profile 
increases greatly at higher frequencies; whereas the more than $\pm$25\degr-longitude 
interval over which significant emission is observed changes hardly at all.  The 
PPAs also reiterate this circumstance clearly;  the discontinuous orthogonal 
polarization mode (hereafter OPM) extends over the full $\gtrsim$50\degr\ width 
of the profiles, whereas the more prominent OPM occupies a more restricted 
longitude range at the lower frequency.  As the PPAs are nearly absolute and the two 
OPMs lie conveniently in the upper and lower halves of the PPA panel, we will 
refer to them as the ``positive'' and ``negative'' polarization mode, respectively.\footnote
{Interestingly, Hobbs \etal's (2005) determination of B1944+17's proper-motion 
direction of 174\degr$\pm$29\degr\ gives the possibility that the fiducial PPA is 
aligned with the star's velocity vector in the sense of Johnston \etal\ (2006) and 
Rankin (2007).  Of course, even if so the double ambiguity of supernova ``kick'' 
orientation and OPM identification makes interpretation impossible at this time.}  
The star's profile has been classified previously (R93a,b) as belonging to the 
conal triple (c{\bf T}) class; looking more closely, however, at the overall L-band 
form, it would be more accurate to regard it as exhibiting a hybrid c{\bf T} and 
conal quadruple (c{\bf Q}) behaviour.  This said, the two profiles are so different 
in form that it is not easy to see how to align them.  Rather, we have used the 
structures of stationary modulation on the leading and trailing edges of the PSs, 
and we note that this tends to align the profile edges but not the peaks.  We will 
come back to considering how these characteristics should be interpreted below.  

\subsection{Non-Random Null Distribution}
In our observations, B1944+17 nulls about 2/3 of the time, somewhat higher 
than the 60\% value given by DCHR, but closer to null percentage reported by 
Rankin, (1986).  The majority of these null pulses can readily be distinguished 
from the bursts; however, there is a small portion of weak pulses that are difficult 
to identify as either nulls or pulses. Interestingly, the distinction between nulls 
and pulses is easier to define at L band, as can be seen in the respective null 
histograms of Fig. \ref{nullhistograms}.  

\begin{figure*}
\begin{center}
\includegraphics[width=160mm]{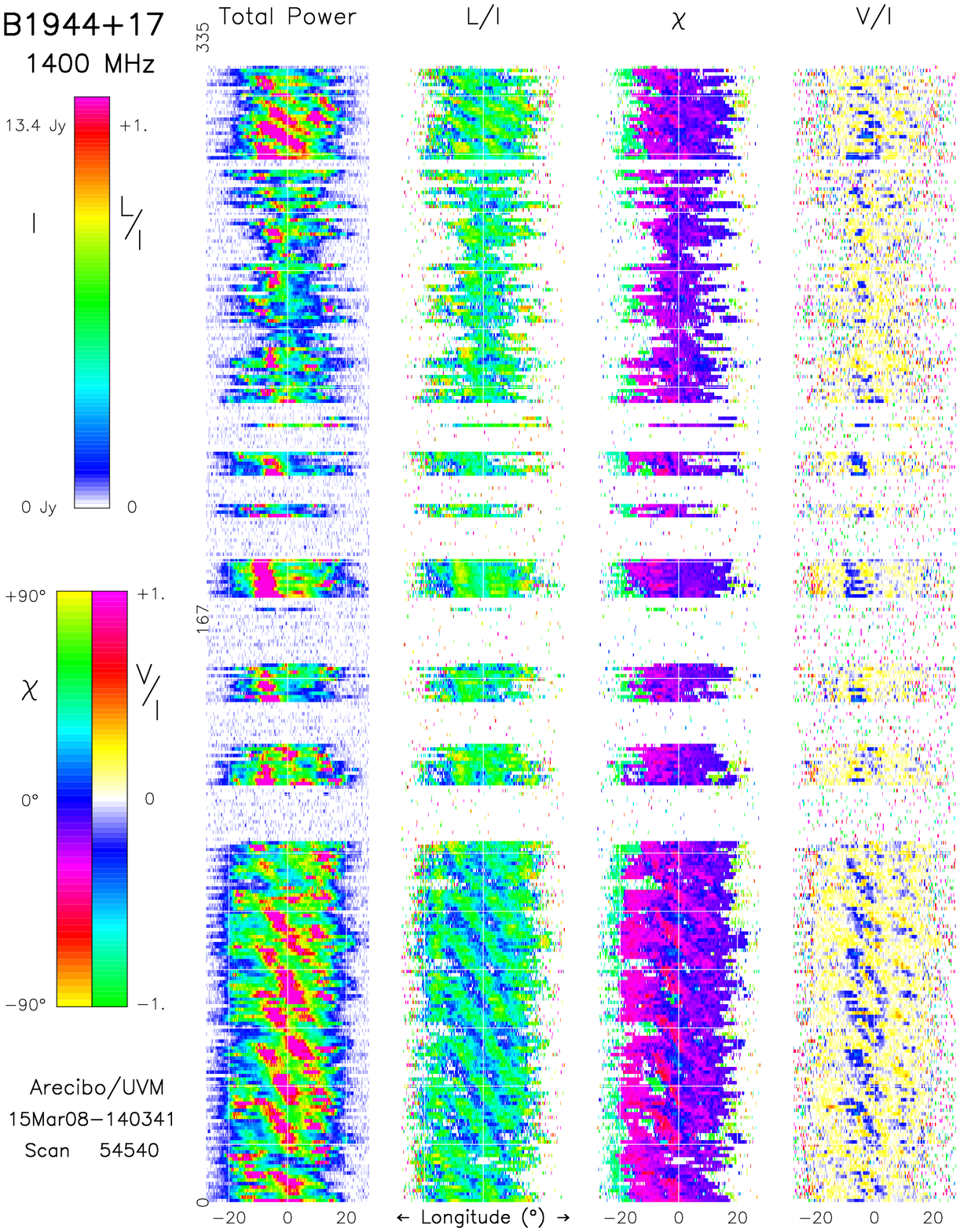}
\caption{Pulse-sequence polarization display showing several 
of the pulsar's PS behaviors, with reordered mode sequences.  The bright and ordered subpulses of mode A 
begin at pulse 1 and last for 110 periods. Mode A is succeeded by mode C, 
separated by a null. Mode C is characterized by roughly stationary subpulses and a 
quasi-periodic cycle of short bursts and nulls. The next 100 pulses (190-290) are 
mode D, the weakest and least structured of the star's modes. Note the weak 
evidence of subpulse structure. Mode D is succeeded by a bright and very well 
defined mode B, separated by a three-period null.  The total 
power $I$, fractional linear $L/I$, PPA $\chi$, and fractional circular polarization  
$V/I$ are colour-coded in each of four columns according to their respective 
scales at the left of the diagram.  Both the background noise level and interference 
level of this observation are exceptionally low with the former effectively 
disappearing into the lowest intensity white portion of the $I$ color scale.}
\label{colourPS}
\end{center}
\end{figure*}

Given that the nulls and pulses cannot be fully distinguished, we can choose 
an intensity threshold that will be conservative and reliable either in selecting 
pulses or nulls, but not both.  In Fig. \ref{nullhistograms}, we have taken the latter 
option---that is, using low thresholds that will tend to slightly underestimate the 
null population.  Then, using this conservative discriminator of nulls, we have 
computed the burst- and null-length histrograms in Figure \ref{burstnullfreq}.  
These show that 1-pulse bursts and nulls have the highest frequency, but we 
see that very long bursts and nulls also occur.  In the 7000-pulse 327-MHz observation, for instance,
a small number of bursts of 40-50 $P_1$ and two of 80-90 $P_1$ were encountered 
alongside the more frequent long nulls ranging up to 300 $P_1$.  Even 
qualitatively we immediately see that the nulls in B1944+17 are distributed within 
the PS in a very non-random manner.  

Recent investigations into pulsar nulling have raised two important new questions 
about their distributions:  a) whether they are randomly distributed (\eg, Redman \& 
Rankin 2009; Rankin \& Wright 2007); and b) whether they are periodic (HR07/09). 
With such a large null fraction, one would expect to see few long sequences in any given 
observation. The tendency of B1944+17's bursts and nulls to clump into sequences of 
roughly 20-100 pulses immediately indicates a non-random distribution.  Application 
of the above RUNS\footnote{The RUNS test for detecting non-randomness returns a 
value greater than 1.96 in absolute value if the data is non-randomly sequenced; see 
Redman \& Rankin (2009).} test to our observations, namely burst and null sequences, 
returns values $\simlt$--60, verifying this conclusion of a non-random ``undermixing''.  
Regarding null periodicity, we find only a suggestion in our observations of a very long 
periodicity---far too long in relation to their total length for it to be significant.

Thus the bursts and nulls of B1944+17 can be regarded as falling into two categories:  
a) short bursts or nulls of some 1-7 $P_1$ that show a roughly random distribution, 
and b) medium to long bursts or nulls ($>$20 $P_1$) that can occasionally persist for 
several hundred pulses and are patently non-random.  We will elaborate further on this 
distinction in \S 5.

\begin{table*}
\begin{center}
\caption{Emission Modes of B1944+17}
\begin{tabular}{c c c c c c c c c }
\hline\hline
Mode&Frequency	&$P_2$			& $P_3$			&Percentage	 		& Number of	& Number of	& Average Burst	& FWHM \\
	&	Band	&	($\degr$)	&	($P_1$)	& of total pulses   		& Bursts		& Drift Bands	& Length		&              \\
\hline
A	& L band		& 12.2$\pm1.0$		& 13.9 $\pm$ 0.6	&8.4$\pm0.05$			& 14			& 25			& 32.5		& 29\\
	& P band		& 13.5$\pm1.0$		& 13.5 $\pm$ 1.1 	&6.4$\pm0.05$			& 11			& 33			& 41.1		& 10\\
B	& L band		& 11.0$\pm1.6$		& 6.5 $\pm$ 1.7   	&3.3$\pm0.05$			& 8			& 9			& 22.0		& 26\\
	& P band		& 8.5$\pm1.7$			& 5.6 $\pm$ 1.8   	&1.4$\pm0.05$			& 8			& 10			& 11.9		& 18\\
C	& L band		& 12.1$\pm1.1$		& n/a                     	&17.9$\pm0.05$		& 118		& n/a			& 8.1			& 28\\
	& P band		& 13.6$\pm0.9$		& n/a				&8.8$\pm0.05$			& 78			& n/a			& 8.0			& 9\\
D	& L band		& n/a					& n/a                     	&4.4$\pm0.05$			& 22			& n/a			& 11.0		& 21\\
	& P band		& n/a					& n/a				&20.6$\pm0.05$		& 119		& n/a			& 12.1		& 12\\
\hline
\label{modes}
\end{tabular}
\end{center}
\end{table*}

\begin{figure*}
	\centering
	\begin{sideways}
	\begin{tabular}{@{}lr@{}lr@{}lr@{}}
	\subfigure[Mode A]
	{\includegraphics[width=50mm]{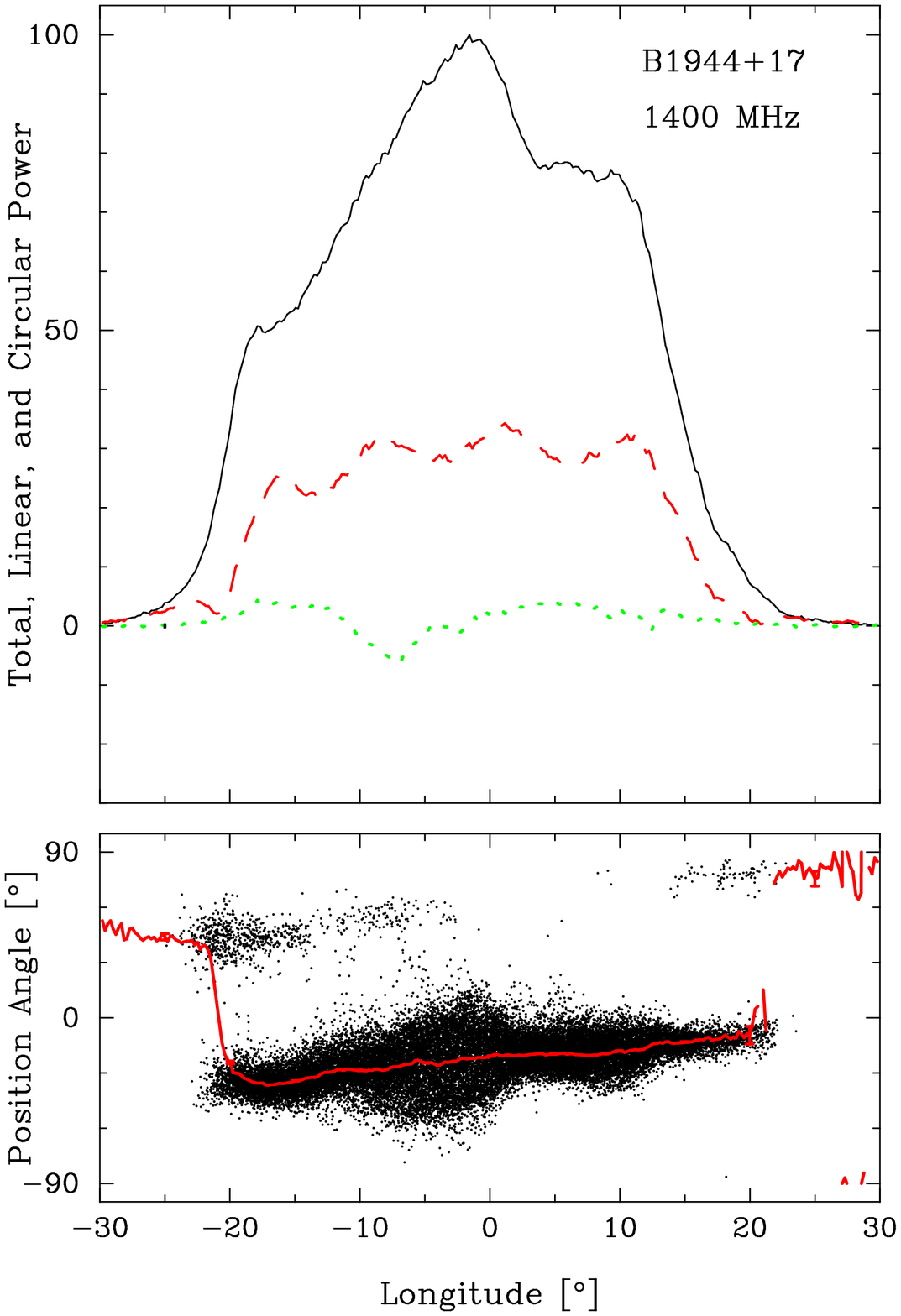}
	\includegraphics[width=50mm]{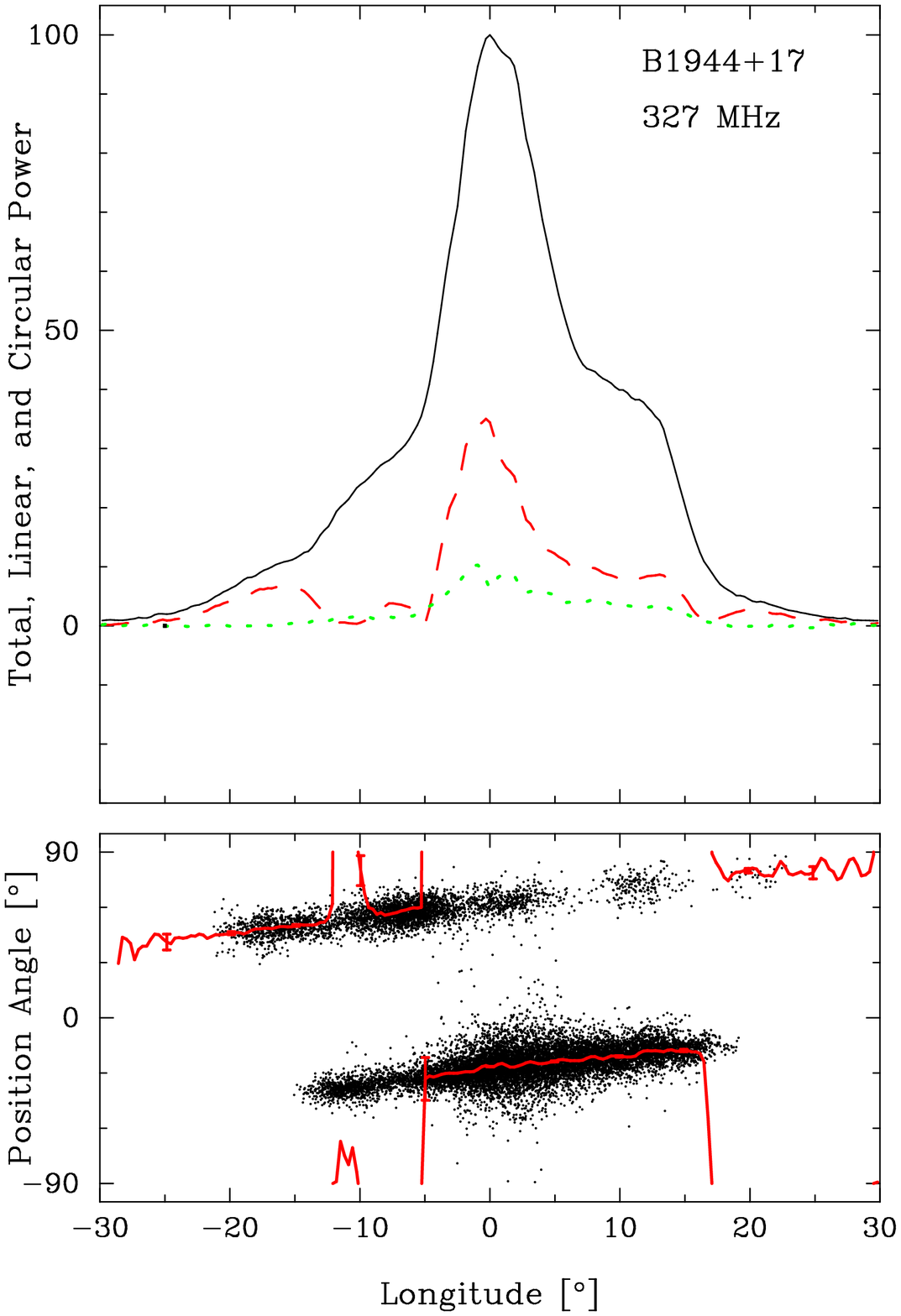}
	\label{modeA}} &
	\subfigure[Mode B]
	{\includegraphics[width=50mm]{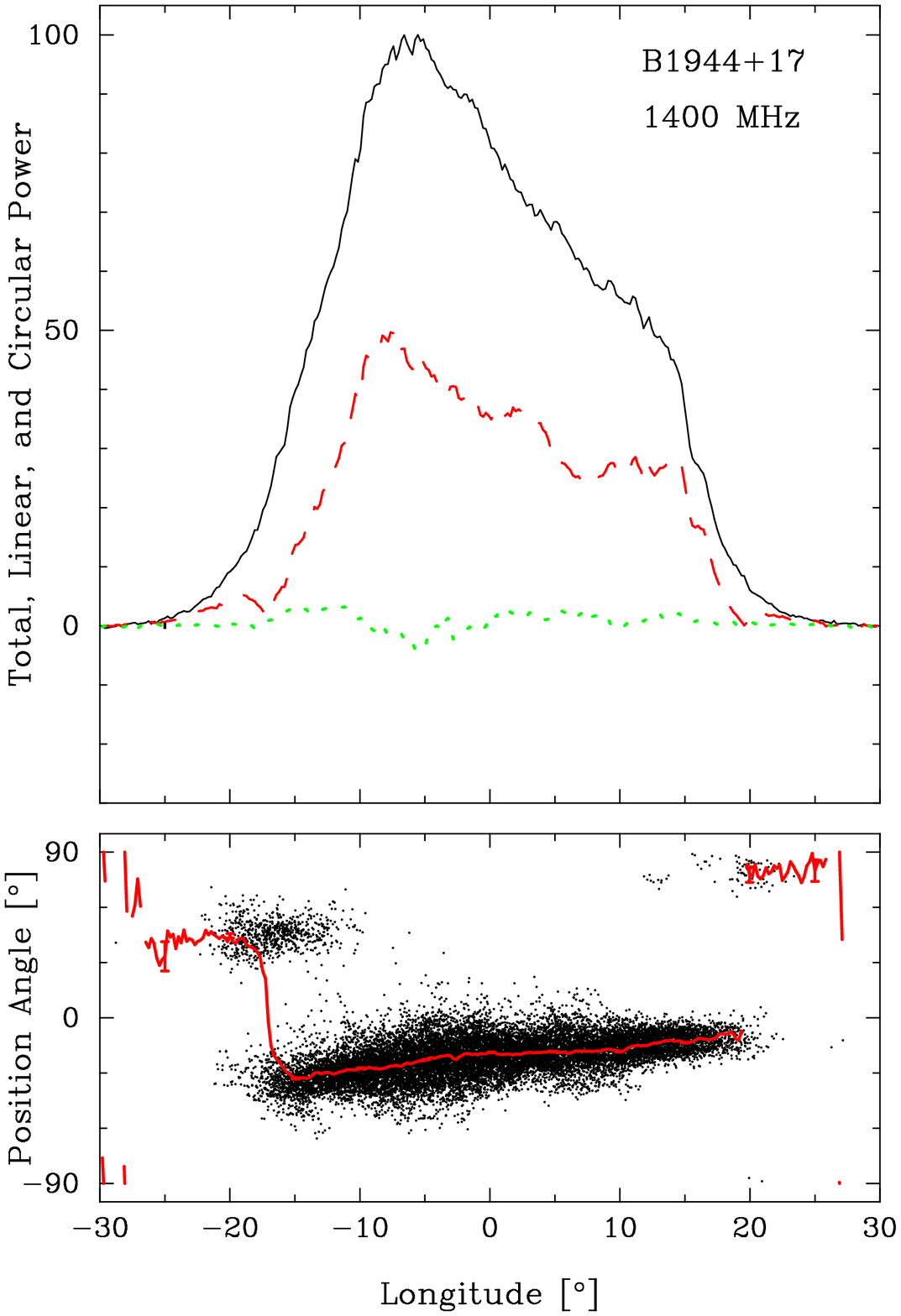}
	\includegraphics[width=50mm]{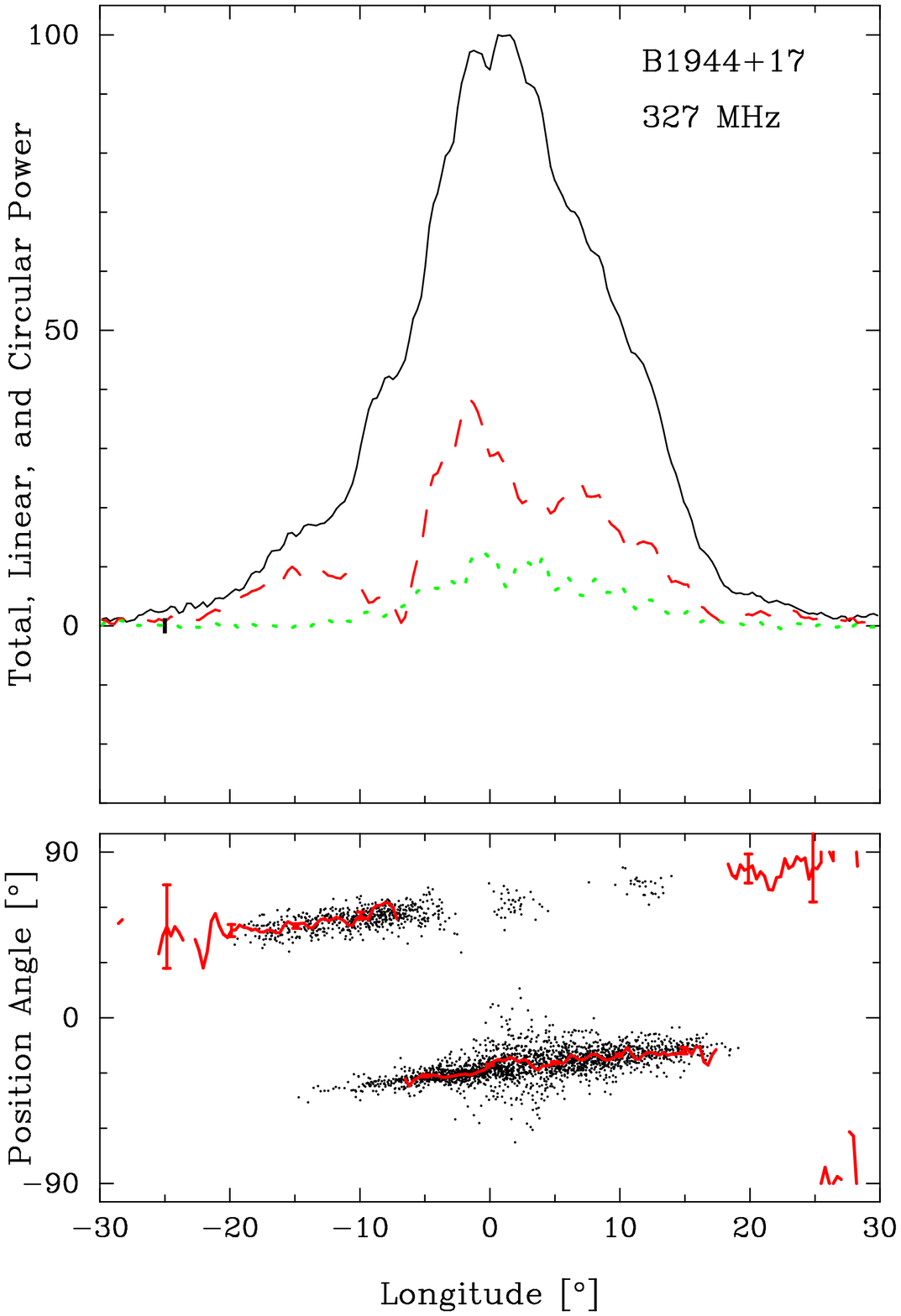}
	\label{modeB}} \\
	\subfigure[Mode C]
	{\includegraphics[width=50mm]{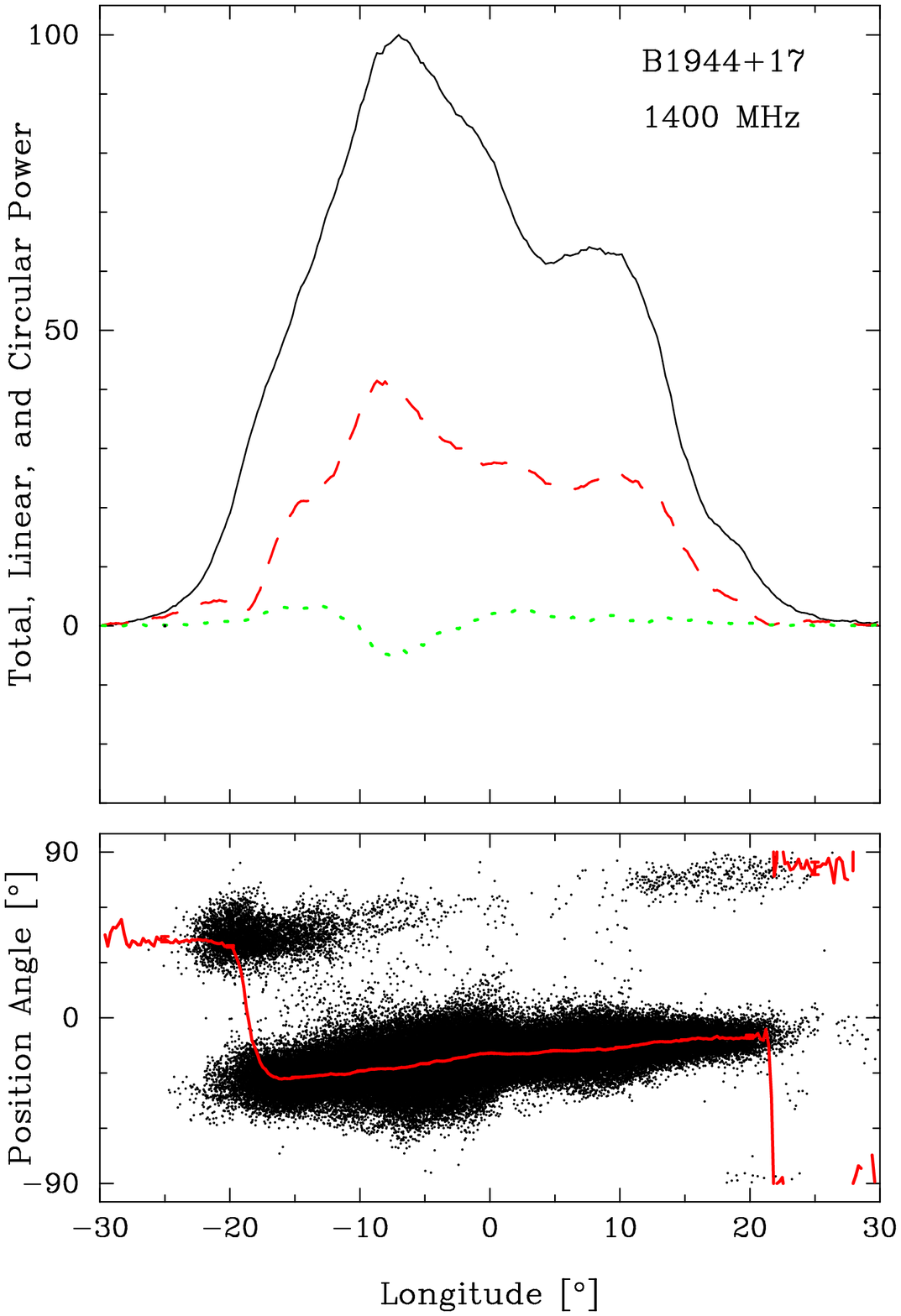}
	\includegraphics[width=50mm]{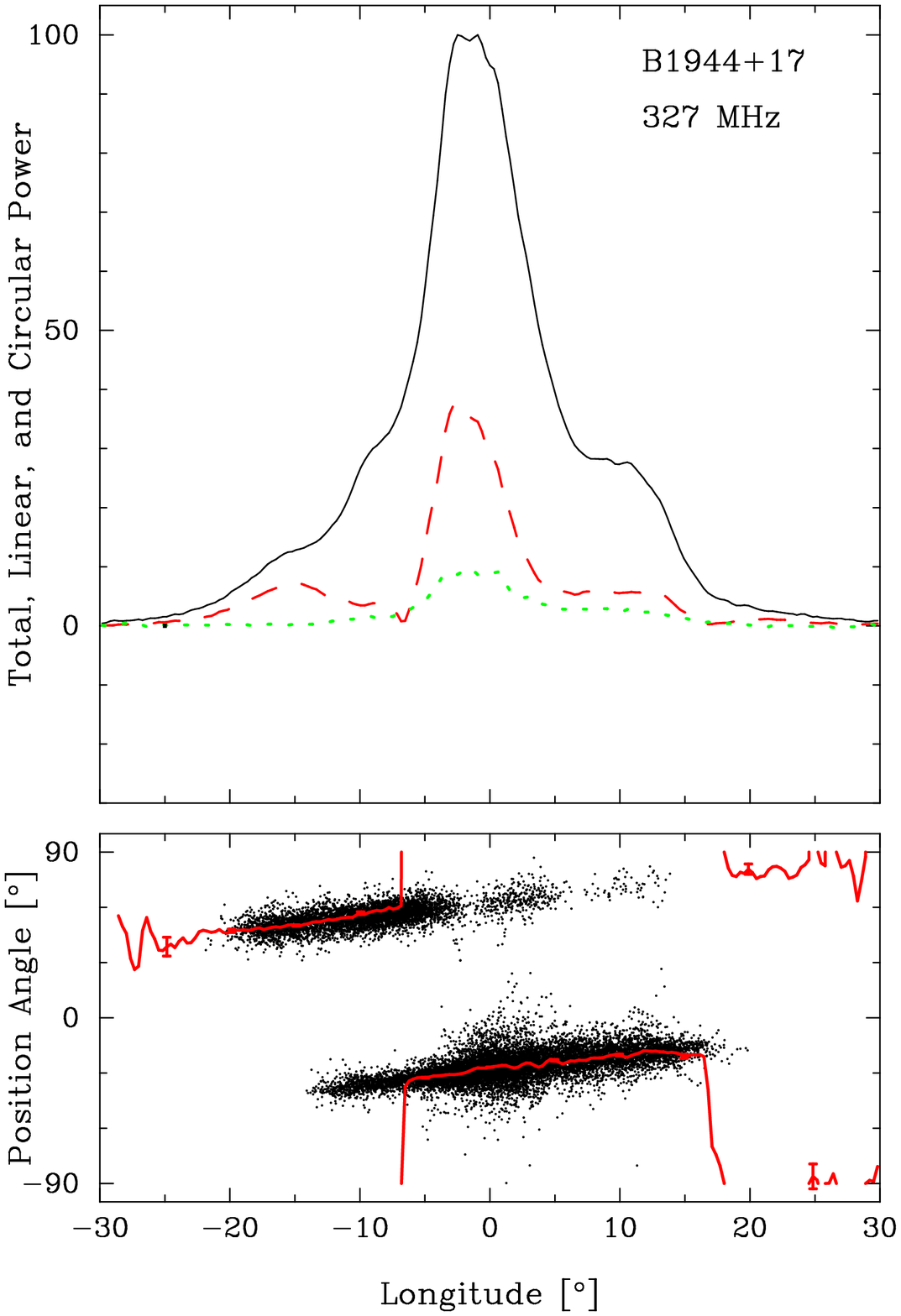}
	\label{modeC}} &
	\subfigure[Mode D]
	{\includegraphics[width=50mm]{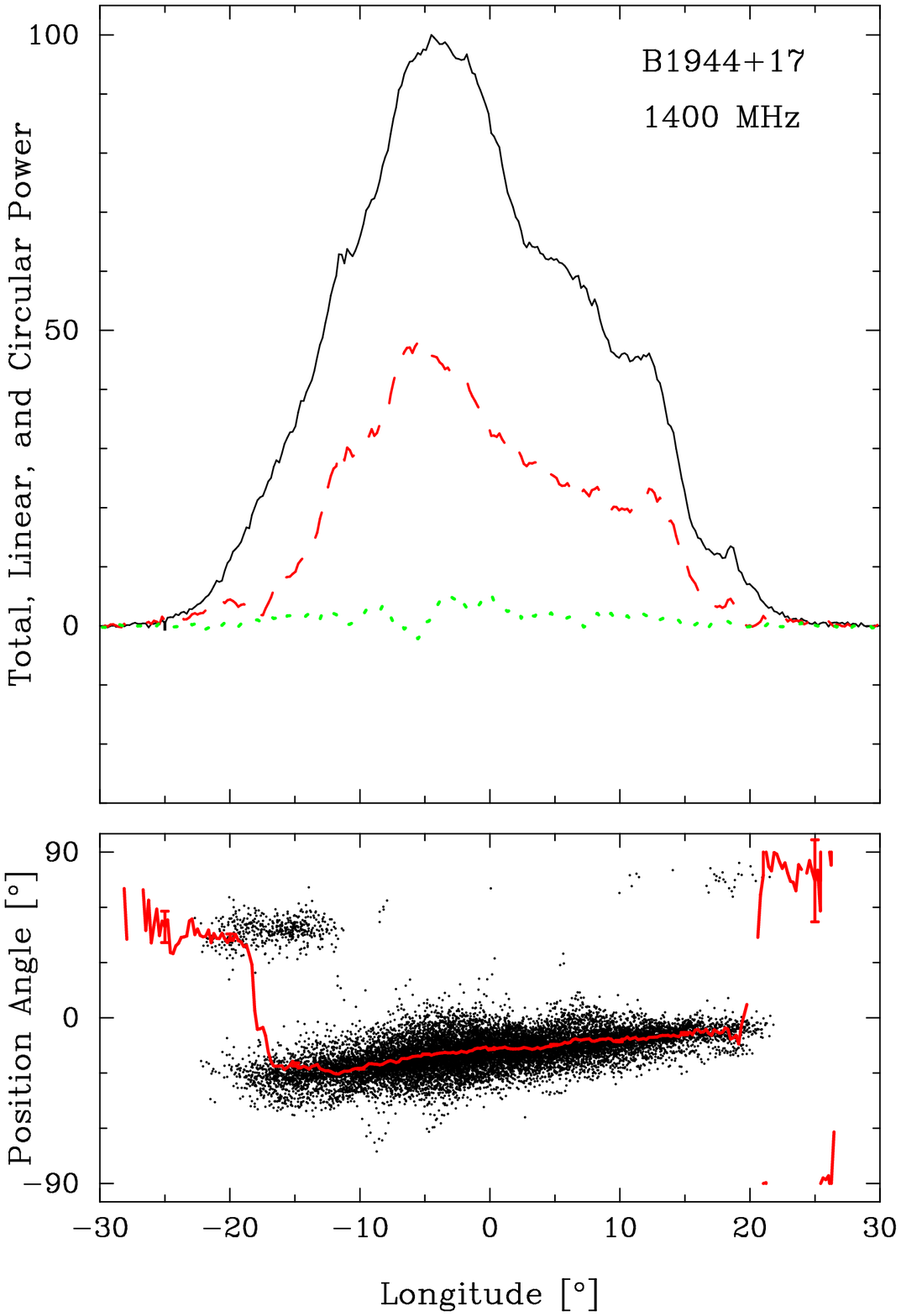}
	\includegraphics[width=50mm]{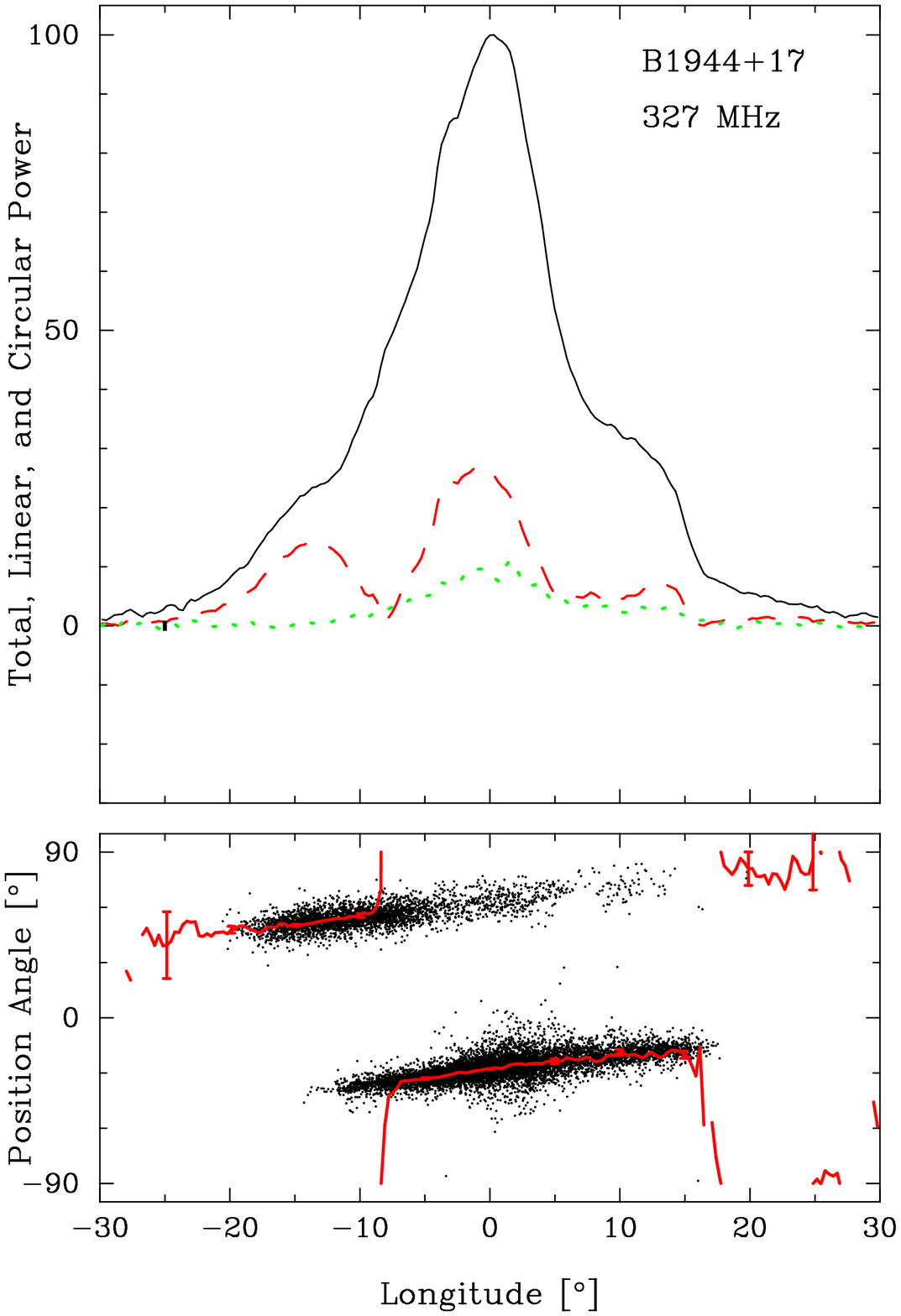}
	\label{modeD}} \\
	\end{tabular}
\end{sideways}
\caption{Partial polarized profiles after Fig. \ref{dottyplots} for the four modes at L 
and P band. The base width within which there is measurable emission is roughly 
constant in all the modes at both frequencies. There is, however, significant variation 
in the FWHMs of the various modes at the two different bands. Most notable is the 
narrowing of the FWHM at P band, clearly displayed in every mode. Also, at L band 
the FWHMs of modes A and C are broader than those of modes B and D.  Conversely, 
at P band the FWHMs of modes B and D are broader than those of modes A and C.  
Numerical values for modal properties are summarized in Table~\ref{modes}.}
\label{modalprofiles}
\end{figure*}

\subsection{Modes of emission}
We here investigate the properties of the four modes identified by DCHR.  
Following their convention, we refer to the three drift modes as A-C, 
and the final burst mode as ``D''. The defining characteristics of modes 
A-D are the same at both P and L band, as are their frequencies of 
occurrence. 

The four modes can be readily distinguished by eye due to their unique subpulse 
structures and intensities, as shown in Figure \ref{colourPS}. The transitions 
between modes occur on a time scale of less than one pulse---\ie, there are 
typically no observable ``transitions'' between modes. \footnote{With two notable 
exceptions, discussed in the null section.} We find that within a sequence of 
$10^3$ pulses there is a high probability of finding at least one occurrence of each 
mode.  It is interesting that this pulsar, which displays an almost overwhelming 
variety of behaviors, is quite reliable in how often it does so. 

As seen in the colour polarization display of Fig. \ref{colourPS}, the star's mode 
changes are usually punctuated by nulls, though there are some combinations of 
mode changes that characteristically occur adjacent to one another. 

\subsubsection{Modal distinctions in subpulse structure}
As modes A and B exhibit subpulse drifting, they can best be characterized by their 
$P_2$ and $P_3$ values, where $P_2$  is defined as the separation of subpulses 
within a period, and $P_3$ is the separation between drift bands at a fixed pulse 
phase. Mode C characteristically displays an organized yet stationary subpulse 
structure. Lastly, we classify those PSs which show no organized subpulse 
structure as mode D; it is worth noting that mode D is significantly weaker than the 
others.

\begin{figure}
	\centering
	\includegraphics[width=72mm,angle=-90.]{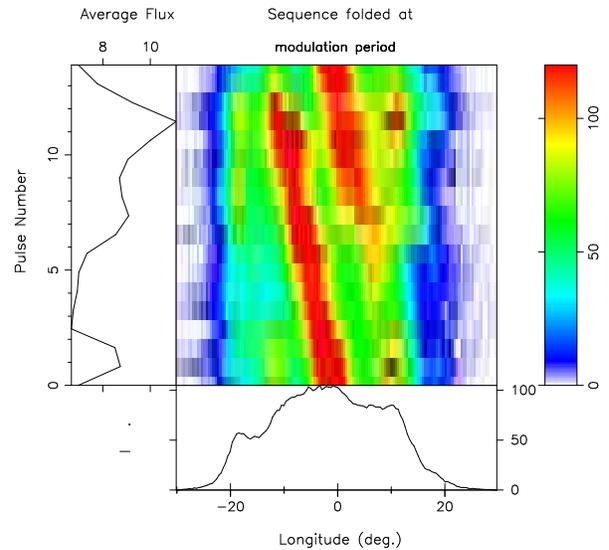}
\caption{A 50-$P_1$ mode-A interval at L band. The PS has been folded at a $P_3$ 
of 14 $P_1$.  The fold length was chosen to demonstrate that the peripheral conal 
outriders, seen here mainly on the leading edge of the pulse window, fluctuate with 
a period equal to mode A's $P_3$.}
\label{modeA_modfold}
\end{figure}

Table~ \ref{modes} gives $P_2$ and $P_3$ values for modes A, B, and 
C.  The A mode is characterized by prominent intervals of remarkably 
precise drifting subpulses---which is paradoxical considering the star's 
otherwise unpredictable and discontinuous behavior.  Mode A is unique 
in its regularity and is marked by its negatively-drifting bands with a roughly 
14-$P_1$ $P_3$. This 14-$P_1$ $P_3$ feature can be seen in an lrf of a 
PS that includes all the modes, indicating its dominance (Weltevrede \etal\, 
2006, 2007).Two bright, central subpulses are usually seen in mode A. 

At L band, weak subpulses on the outer edges of the profile turn on and 
off with a period that is comparable to mode A's $P_3$; see Figure~\ref{modeA_modfold}. 
Mode A always appears in bursts having durations of more than 15 periods; 
however, usually these bursts are even longer, typically some 60 -- 100 
$P_1$---and, remarkably, these A-mode intervals are very rarely interrupted 
by nulls.  

The drifting subpulses of mode B are visibly less ordered than those of mode A; 
however they are clearly structured and negatively-drifting. $P_3$ is approximately 
half that of mode A at both L and P band; although due to its irregularities 
these values cannot be measured with the same degree of precision as for mode 
A.  The B mode characteristically persists for less than 25 $P_1$.

The third ``drift'' mode, C, displays three roughly stationary subpulse drift bands 
($P_3$/$P_2\approx 0$), with the components' relative intensities being variable. 
Mode C manifests itself in a complex variety of ways:  most often the intensities 
of the three components are approximately equal to each another; see 
Fig.~\ref{colourPS}.  However, one or two of the components intermittently 
either turns off or notably weakens relative to the other two; every combination 
of the three constituent components was observed. The $P_2$ value for mode 
C is similar to that of mode A, taking into account the presence or absence of 
its three constituent features.

It is possible that the nearly vertical drift bands in mode C result from a near 
stoppage of carousel rotation; it is also possible that this is an effect of aliasing.
Because of the other similarities between 
modes A and C, it is likely that this ``stopping'' happens during what is 
otherwise known as mode A. The difference between modes A and C is 
therefore only the carousel motion, not the fundamental subpulse structure. 
This dynamic accounts for the varying number of subpulses observed in 
mode C. In mode A, as the subpulses drift across the pulse window, 
anywhere between 1 and 3 may be seen depending on the modulation 
phase---\ie, see Fig.~\ref{modeA_modfold}. 

In mode C short bursts and nulls alternate quasiperiodically with each burst 
or null lasting some 10 $P_1$ or so, and switching back and forth up to 10 
times. The length of these segments is very comparable to the $P_3$ of 
mode A. 

Deich \etal\ first referred to mode D as ``chaotic''---that is, displaying little 
perceptible order in its subpulses. Our investigation has uncovered a slightly 
different story for mode D. The mode-D emission does not usually span the 
entire pulse window, see Fig. \ref{colourPS}. While its sparse subpulses are 
overshadowed by the bright and ordered ones of modes A, B and C, they 
appear to consistently have an underlying characteristic structure.  Mode D's brightest 
subpulses appear approximately every 5--10 $P_1$ on the leading edge 
of its substantially narrower emission window. This withdrawal of emission 
in the profile wings is unique to mode D.

\subsubsection{Modal distinctions in total and linear power distributions}
Figure \ref{modalprofiles} gives partial polarized profiles for each of the four 
modes at L and P band. The profiles have different total power and total 
linear forms in the different modes. Most evident are their different modal 
widths in each of the bands. 

Mode A has the broadest modal profiles; see Figure \ref{modeA}. Its L-band 
profile displays a nearly uniform distribution of linear power across the pulse 
window, showing only small dips that correspond to the subpulse separation. 
Its leading edge is marked by a distinct shoulder at both L and P band, also 
visible in the linear power, a feature hardly seen in any of the other modes at 
L band. The bright central subpulses in mode A display a linear power 
distribution that indicates there are in fact two distinct features that lead the 
central peak. They can be identified at both bands, though the second one is 
weaker at P band, nevertheless distinguishing itself as a unique feature. Also 
in both bands, the trailing subpulse features are brighter and more clearly 
defined than the leading ones. 

The mode-B profiles display a slightly narrower FWHM than in mode A; see 
Table~\ref{modes} and Fig. \ref{modes}. The leading feature in mode A's profile 
is clearly visible in mode B at P band but is very weak at L band; see Figure~\ref{modeB}. 
At L band, the central peak of mode B occurs $\sim$6\degr\ earlier than in 
mode A; at P band the peaks of the two modes occur in the same location. 
The trailing feature in mode B is poorly defined, however this is likely due to 
the dense drift bands blurring the intensity distribution between the leading 
and trailing features, rather than a decrease of subpulse intensity.

Mode C displays a FWHM width nearly as broad as that of mode A; see 
Table~\ref{modes}. There is more evidence for the leading feature in the 
linear power distribution of mode C than in mode B. As in mode B, the central 
peak in mode C at L band is shifted 6\degr\ earlier relative to the peak in 
mode A, whereas at P band they occur at the same longitude. The trailing 
component is very well defined in mode C in the profiles of both bands. 

 \begin{figure}
	\centering
	\includegraphics[width=72mm,angle=-90.]{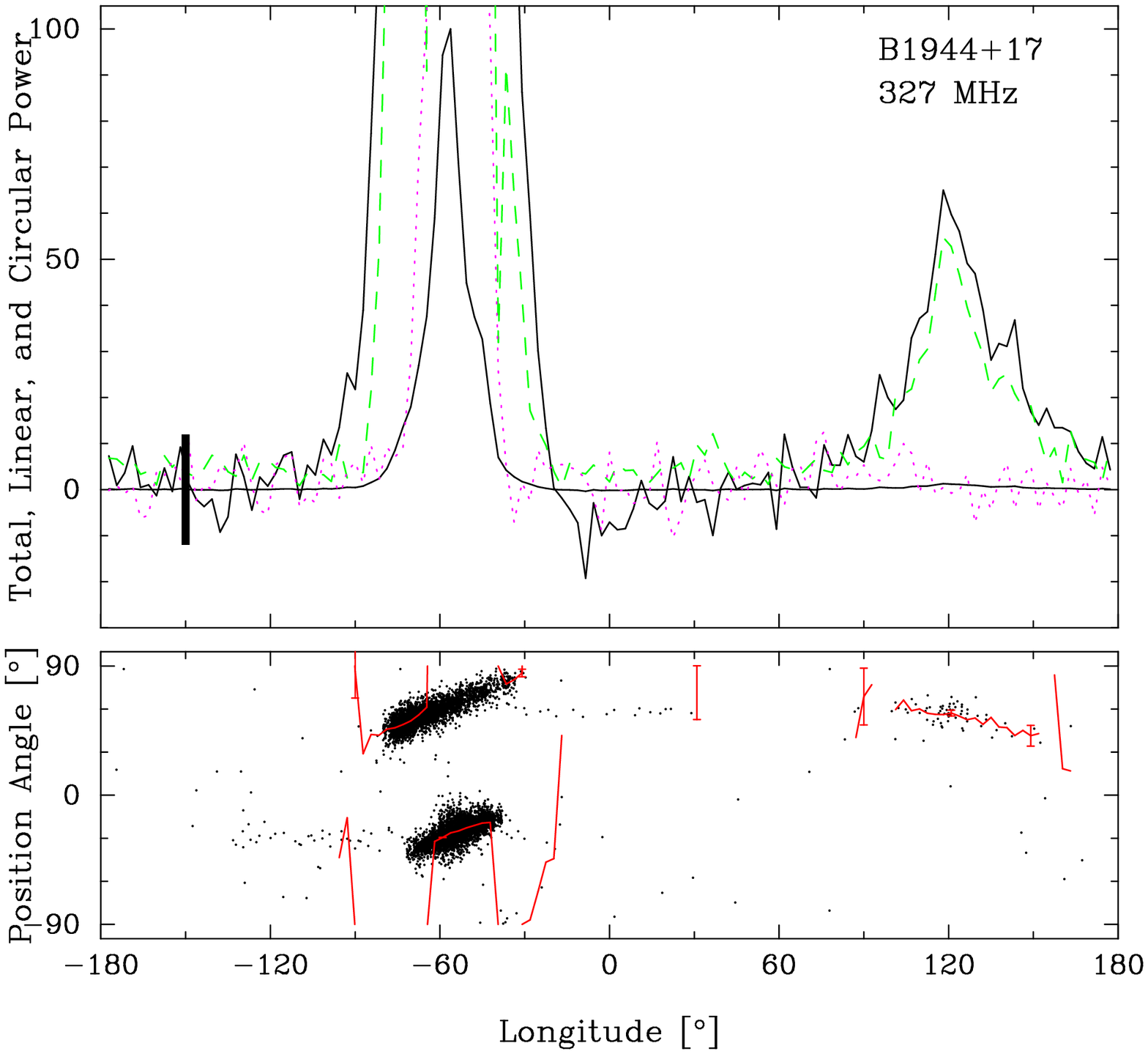}
	\includegraphics[width=72mm,angle=-90.]{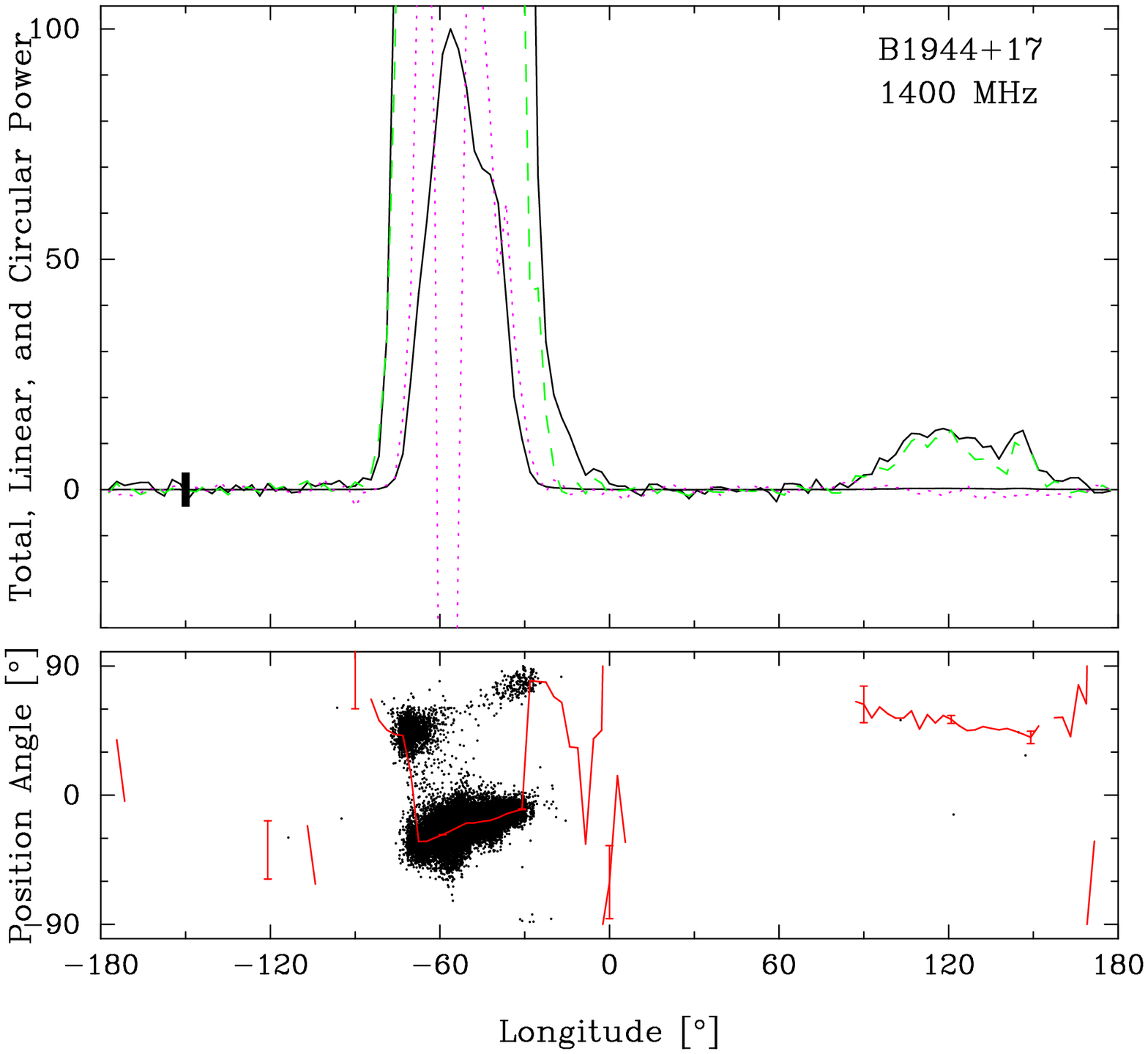}
\caption{Full period polarization profiles (after Fig. \ref{dottyplots}) of the P- (upper) 
and L-band (lower) partial PSs containing the pulses (falling above the thresholds 
in Fig. \ref{nullhistograms},  numbering 2403 and 1803 pulses, respectively), 
not the nulls.  Here, the total power is 
shown at full scale and then Stokes $I$, $L$ and $V$ are replotted at 50X 
scale in order to reveal the structure of the interpulse; otherwise the displays 
are identical to those in Fig. \ref{dottyplots}.  Only 128 longitude bins were 
used across the full profile so as to maximize the S/N of each sample.  Note 
the virtually complete IP linear polarization and the similarity in half-power 
widths between the IPs in the two bands.  Note also that the centers of the 
PPA traverses of the IP and MP (positive OPMs) both fall at about +60\degr\ 
and have opposite slopes.}
\label{inflatedprofiles}
\end{figure}

Mode D displays a significantly narrower profile at L band compared to the 
other modes. As in mode B, its leading component is very poorly defined at 
L band, though there is some evidence for it in the linear power distribution. 
At L band, the central peak of mode D is shifted approximately 4\degr\ compared 
to that of mode A. At L band, the trailing feature is slightly more distinct in 
mode D than in mode B, though it remains weak. The trailing feature at P band 
is well defined.

\subsubsection{Mode changes}
Mode changes are often punctuated by nulls, with two distinct exceptions:  First, 
a small proportion of mode changes are characteristically not interrupted by a null. 
We find that mode D often precedes modes A and B and sometimes succeeds 
mode B continuously. Second, nulls do not necessarily result in mode changes. 
Mode C (discussed below) characteristically displays a pattern in which the 
emission turns on and off in segments of short 10-15 $P_1$ bursts and nulls. 
Short nulls are associated both with this mode-C emission pattern as well as 
with transitions between modes A, B, C and D.

\subsection{Main Pulse -- Interpulse Relationship}
We have used the conservative thresholds of 0.28 and 0.20 $<I>$ to build 
partial P- and L-band PSs of the pulses and nulls.  Full period polarization profiles 
corresponding to the pulses (not the nulls) are shown in Figure \ref{inflatedprofiles} 
and are similar to the MP profiles in Fig. \ref{dottyplots} except that Stokes $I$, 
$L$ and $V$ are replotted at 50X scale to reveal the structure of the IP.  Here we 
see clearly for the first time that the B1944+17 IP is almost fully linearly polarized 
(and has negligible circular), that its half-power width is very comparable at the 
two frequencies, and that its IP PPA traverse is centered at virtually the same 
angle as the MP positive-mode, but with a negative slope.   In the 327-MHz profile, 
the discrete PPA dots under the IP profile show that some IP samples are strong 
enough to define their linear polarization; thus the IP is not uniformly weak, but 
exhibits a range of intensities.  We can also confirm that  the IP-to-MP intensity 
ratio ($S_{IP}/S_{MP}$) decreases with frequency, being about 1\% at P and 
some 0.2\% at L band.  HF86 reported values of about 3\% at 430 MHz and 
0.3\% at L band. Notably, they found that most pulsars with IPs show a similar 
decrease as frequency increases. More perplexing was HF86's finding that the 
MP-to-IP spacing ($\Delta\phi_{IP-MP}$) decreases by some 10\degr\ between 
P and L band, and given the drastic changes in MP form it is not fully clear how 
such a change could even be consistently measured.\footnote{Thanks to Tim 
Hankins, we have been able to examine the interpulse-discovery observations 
that were reported in HF86, and we can see how their interpretation of a 
frequency-dependent $\Delta\phi_{IP-MP}$ arose.  At that time, it could not have 
been realized that the unusual shape changes of B1944+17's MP make it nearly 
impossible to correctly align profiles of different frequencies.}

Using such partial PSs for the pulses (but now at both the restricted and full  
longitude resolutions), we also computed correlation functions that included 
the longitude ranges of both the MP and IP.  These showed no significant 
($>$3 standard deviations in the off-pulse noise) level of correlation at lags 
of either 0 or $\pm1$ pulse.  

We also computed partial profiles for the null partial PSs, having lengths of 4635 
and 3667 at P and L band, respectively.  In these, we were able to find neither 
significant total power nor any correlated PPAs that might indicate weak linear 
polarization in the baseline regions.  This circumstance argues very strongly 
that the IP either nulls, or remains very inactive, during MP nulls. 

\section{Emission Geometry of the Main Pulse and Interpulse}
\label{sec3:III}
At various stages of our analyses we had indications that the baseline regions 
of the full period profiles were significantly linearly polarized.  In order to explore 
this further, we reprocessed our observations in a manner such that no baseline 
level was subtracted from Stokes $Q$ and $U$.  We then found a significant 
level of linear polarization in both regions between the MP and IP corresponding 
to some 0.25\% of the MP peak at P and 0.8\% at L band.  In both cases the PPAs 
associated with this baseline linear polarization are nearly flat, suggesting that 
only one Stokes parameter is well defined, therefore we have concluded that this 
baseline polarization cannot be well measured with our polarimeter configurations.\footnote
{The L-band system consists of dual linears and the P band dual linears with a 
circular hybrid between the feed and the cal injection; thus in neither system are 
Stokes parameters $Q$ and $U$ determined fully by correlation of the receiver 
voltages.}  Again, we stress that this baseline linear power is associated with the 
pulses and not the nulls!  HF86 also allude to a ``bridge'' of total power emission 
between the star's MP and IP (see their fig. 2d and the associated discussion).

Regardless of whether the star is classified as a member of the c{\bf T} 
or c{\bf Q} class, it can be said with confidence that our sightline crosses 
two concentric emission cones.  The weaker cone is encountered on 
the profile edges and exhibits similar properties and dimensions at 
the two frequencies.  It is the stronger central emission that shows an 
``unresolved double'' pattern above 1 GHz and a more ``single'' form at 
meter wavelengths.  Two aspects of this profile's evolution are quite 
unusual (\eg, R93a,b):  a) that the central emission pattern narrows 
(shows a smaller equivalent width) at lower frequencies, and b) that 
the ``outriding'' component pair retains essentially the same dimensions 
between the two bands.  Generally, we encounter outer cones on 
profile edges, and these show substantial ``spreading'' with wavelength; 
whereas inner cones are seen inside the outer ones and show little 
spectral variation in their dimensions.  In B1944+17 these usual 
expectations seem reversed; how could this occur?

\subsection{Assembling the Relevant Evidence}
Careful reference to the total power profiles of Fig. \ref{inflatedprofiles} 
shows that the MP and IP are both rather broad---as is expected for 
a small value of $\alpha$---with emission extending over most or all 
of 60\degr\ in each case.  Moreover, (HF86 notwithstanding), the MP to 
IP spacing is close to half the rotation period.  Given the complex and 
changing form of the MP in particular, it is difficult to be more precise 
than this.  

Several aspects of the linear polarization are also important to note:  First, the 
absolute PPAs associated with the central longitudes of the MP and IP are 
nearly identical---allowing for the 90\degr\ OPM ``jumps---arguing that the 
respective emission regions are either conjunct or in opposition along a given 
magnetic longitude.  Second, the PPA traverses under the MP and IP have 
opposite senses.  At both frequencies, 90\degr\ OPM-dominance 
``jumps'' occur on both the leading and trailing edges of the {\bf MP} profiles. 

Third (see Fig. \ref{dottyplots}), the MP PPA traverse is unusually 
shallow.  Its sweep rate $R$ ($=\Delta\chi / \Delta\phi$) is only some 
{\bf +0.75}\degr/\degr\ and remarkably linear across the bodies of both the 
P and L-band profiles.  The shallow PPA traverse indicates that the 
magnetic colatitude $\alpha$ and sightline-impact angle $\beta$ are 
similar and both small (\ie, the sweep rate $|R|$=$\sin{\alpha}/\sin{\beta}$).  
Interestingly, the negative sweep rate of the IP has an even shallower 
slope.  

Finally, the interpulse emission is of an entirely different character than 
that of the MP.  The MP has a complex structure of regions with different 
dynamics and OPM activity, whereas the IP is more nearly unimodal in 
form (though with low persistent peaks) and fully linearly polarized.  In 
contrast to the MP, it displays a continuous, smooth PPA traverse---\ie, 
it does not show any ``90\degr\ jumps'' as the MP does.  Moreover, 
the IP emission seems uncorrelated with that of the MP. These MP and 
IP properties provide crucial information about the emission geometry 
as we will see below.  

\subsection{Building A Geometrical Model}
While B1944+17's IP has been known for more than 20 years (HF86), its 
polarization properties are measured here for the first time.  Therefore, all 
previous efforts to understand the star's basic geometry have been based 
on its MP properties alone.  HF86 speculated about whether the IP reflected 
a single or two-pole configuration, but made no attempt at a quantitative 
model.  The star's unusually shallow $R$ value presents a major difficulty 
for any model.  Indeed, the only published model (R93a,b) views the pulsar's 
MP profile as c{\bf T} and obtains reasonable dimensions for the inner and 
outer cones---but only by assuming that the apparent $R$ value was somehow 
too flat.  The actual sweep rate of 0.75\degr/\degr\ implies that $\alpha$ 
will be even smaller than $\beta$.  This also suggests strongly that the observer's 
sightline to B1944+17 largely remains inside its emission cones!  A current 
PPA fit to our same 327-MHz observations that includes the IP (Mitra \& Rankin 
2010; see their fig. A8) gives highly correlated (98\%) $\alpha$ and $\beta$ 
values of 2.8 and 4.0\degr, respectively, with large errors ($\pm$15\degr).  
Clearly, somewhat different values of $\alpha$ and $\beta$ that maintained 
$R$ would also fit the PPA traverse well.

 \begin{figure}
	\centering
	\includegraphics[width=82mm,angle=-90.]{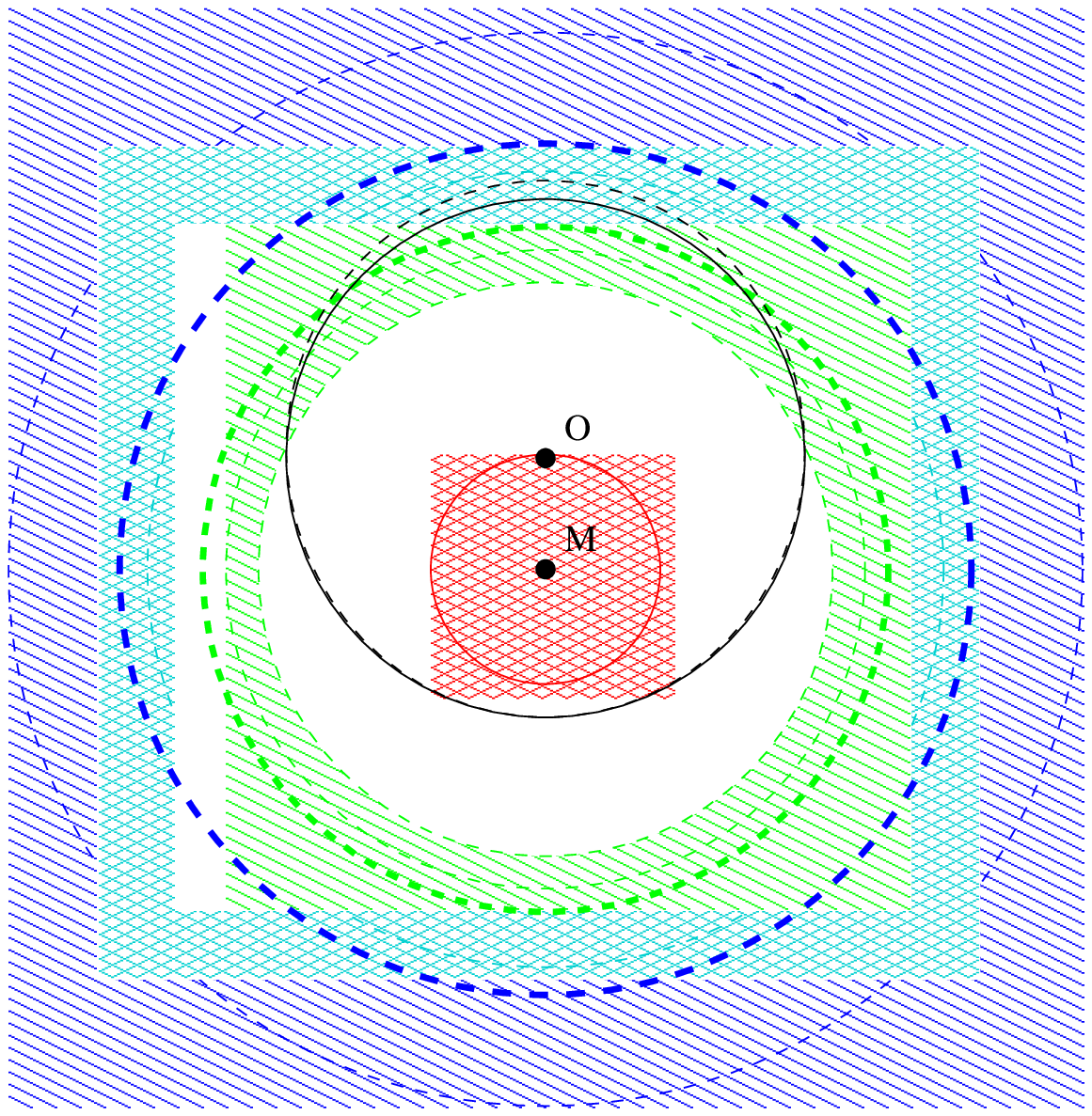}
\caption{Diagram showing the apparent sightline geometry of pulsar 
B1944+17.  The outer cone is shown in blue, the inner cone in green, 
and their overlap region in cyan.  The core beam is then shown in red.  
The radial peaks of the respective cones are shown by thick blue and 
green dashed lines.  The edges of the beams show their half power 
points (see text).  The magnetic axis is marked with an ``M'' and the 
rotation axis with an ``O''.  As we know little for B1944+17 about the 
frequency dependence of the beam dimensions, we have used their (R93a,b) 
nominal 1-GHz values---e.g., 8.7\degr\ for the outer beam's outer 3-db 
radius (see text).  The observer's sightline orbit then makes a small 
circle around the rotation axis such that it touches both cones and 
the periphery of the core beam.  The beam pattern has a larger angular size at the 
lower frequency, so we show the {\it relative} size of the sightline 
orbit within the beam pattern.  The L and P-band sightline orbits} are then 
indicated by dashed and solid black curves, respectively.
\label{beamgeometry}
\end{figure}

The half- or equivalent widths of the MP and IP are roughly equal and 
occupy opposing $\sim$60\degr\ or 1/6 sections of the star's rotation 
cycle.  Moreover, the similar PPA ranges and opposite slopes of the 
MP and IP traverses argue that our sightline cuts these regions within 
the same range of magnetic longitude (modulo 180\degr) but at different colatitudes.  This 
symmetry is consistent both with an orthogonal rotator model (in which 
the MP and IP emission comes from the two respective poles) as well 
as a single pole model (wherein both MP and IP are emitted within a 
single polar region).  Notably, the extreme shallowness of $R$ and the 
small values of $\alpha$ and $\beta$ are more consistent with the latter 
configuration. 

The very small $\alpha$ and $\beta$ values indicate that the pulsar is a      
nearly aligned rotator with the Earth positioned almost directly above its 
``nearer'' polar cap of emission, similar to the single pole IP model first  
proposed by Gil in 1985.  According to this model, the MP is a result of our 
sightline first crossing the inner cone of emission obliquely, then making 
a tangential cut through the inner/outer cone overlap region, and 
finally recrossing the inner cone symmetrically.  Such an ``inside out'' 
sightline traverse accounts for the stability of the outer parts of the MP 
profile as well as the unusual frequency dependence of the middle region.  
Moreover, the weak IP feature can apparently be understood as a grazing 
encounter of the sightline with the far ``skirts'' of the core beam---such 
that the MP and IP are centered on opposing field lines that are cut in 
the same directions.  

Finally, we can assemble the elements of a quantitative model for the 
emission geometry of B1944+17.  Following R93a,b we know that the outer 
and inner conal beams have particular dimensions at 1 GHz, here specified 
in terms of the outside half-power points (which for a 440-ms pulsar) are 
8.7\degr\ (=5.75\degr$P_1^{-1/2}$) and 6.5\degr\ (=4.33\degr$P_1^{-1/2}$),
respectively.  A similar model is available for the peaks of conal beams 
(Gil \etal\ 1993) such that these fall at 6.9\degr\ (=4.6\degr$P_1^{-1/2}$) 
and 5.6\degr\ (=3.7\degr$P_1^{-1/2}$).  No such study has determined 
the inner half-power dimensions of conal beams, but if we assume that 
they are radially symmetric, we can estimate their inner dimensions from 
the above data.  These values are then 5.2\degr\ (=3.45\degr$P_1^{-1/2}$) 
and 4.5\degr\ (=2.97\degr$P_1^{-1/2}$) as above, respectively.  These conal 
beam characteristics are illustrated in Figure \ref{beamgeometry}:  the outer 
and inner cones are hashed in blue and green, respectively, up to their 
half-power levels, and their peaks are indicated by heavy colored dashed 
lines.  Note that the two beams overlap significantly (as they often do in 
actual profiles) and this overlap region is hashed in cyan.  Finally, the core 
beam is shown in red hashing centered on the magnetic axis (``M'') out to 
its 1.8\degr\ (=2.45/2$P_1^{1/2}$) half-power point.  


The respective L ({\bf dashed}) and P ({\bf solid}) sightline trajectories are 
then fitted into the above radiation-beam geometry as indicated in 
Fig.~\ref{beamgeometry} by the black sightline curves centered on 
the rotation axis {\bf (``O'').  The magnetic colatitude $\alpha$ is about 
1.8\degr, the sightline impact angle $\beta$ about 2.4\degr, making the 
sightline circle radius just over 4\degr.}  Of course, the angular radius of 
the sightline circle does not vary with frequency.  Given, however, 
that the B1944+17 profiles provide very little information about the ``conal 
spreading'' at meter wavelengths, we have chosen to illustrate the pulsar's 
geometry using the well determined 1-GHz dimensions of the beam structure.  
In relative terms then, the sightline at L band extends well past the radial 
maximum point of the {\bf inner} cone, such that it exhibits a c{\bf Q} structure; 
whereas the P-band traverse falls just short of this point and has a 
c{\bf T} form.  In both cases the sightline encounters the core producing 
the IP.  

The multiple lines of evidence that our analyses of this pulsar provide 
point strongly to the unusual emission geometry discussed above. Were 
reliable baseline polarimetry available, the sightline traverse could be 
modeled in considerable detail and the actual dimensions of the several 
beams estimated through fitting.  Unfortunately, this work remains beyond 
the scope of this paper.  Nonetheless, we believe that the circumstances 
responsible for the B1944+17 interpulse are largely understood---that both 
the MP and IP are emitted by a single pole and that the IP very likely 
represents weak core emission.

\section{How Can We Understand B1944+17's Nulls?}
Pulsar B1944+17 has been most famous for its remarkably long null intervals, 
which contrast so beautifully with its intense and well organized burst sequences. 
The star is in the null state some two-thirds of the time, on average. Notably, this value
 is the same for both the main pulse and the interpulse.

Since Backer's first documentation of pulsar nulling in 1970, the principal question 
surrounding nulls has been one of causality. Here we discuss and distinguish 
between nulls that are likely caused by two entirely different mechanisms: empty 
sightline traverses across a rotating carousel, and cessations of emission. We find 
that the two null mechanisms can largely be distinguished by their characteristic 
lengths, and we will therefore refer to them as such: the former being short nulls 
($\lesssim$7 $P_1$), and the latter long ones ($\ll7P_1$).  Note that this 7 $P_1$ 
boundary falls on the tail end of the random distribution of null lengths in Fig.~\ref{Nulls}, 
whereas the long nulls quickly become non-random.

\begin{figure}
\centering
{\includegraphics[width=80mm]{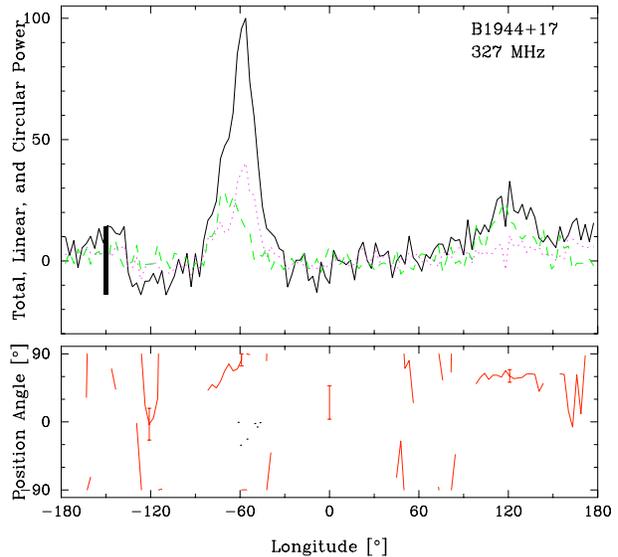}
\label{longnullprofile}}
\caption{The short null profile---\ie, the sum of all 296 P-band pulses which 
participate in nulls less than 8 periods in length.  An intensity threshold of 0.28 
$<$$I$$>$ was used to differentiate between nulls and bursts.  Then, those 
pulses in the short null sequence which still displayed visible power (appeared 
to possibly be weak bursts) were removed by hand. The remaining weak MP 
and IP profile corresponds to an intensity threshold of 0.18 $<$$I$$>$, and 
indicates that there is a weak intensity signal (not detectable on a single pulse 
basis) present in the pulses which participate in short nulls.}
\label{nullprofiles}
\end{figure}

\subsection{Null Analysis}
It is important to note that the distinction between short and long nulls is not a 
clear one. In this investigation, the numerical boundary was set by averaging the 
total power in a series of null sequences, with varying upper bounds in null length, 
then selecting the short versus long null boundary to be where the average power 
dropped off.  This procedure makes no claim of being exact.  We are primarily 
interested in the possibility that B1944+17 exhibits two distinct types of nulls, 
rather than in defining their features precisely.

\subsubsection{Short Nulls}
Short nulls can be found in virtually any context of this star's emission---that is, 
they are not unique to a particular mode, nor do they only occur between mode 
changes.  The short null average profile was constructed by summing all the 
nulls (falling below the established threshold, 0.28 $<$$I$$>$ at P band) which 
participated in null sequences $\leq$7 $P_1$ in length (after removing a few 
with obvious ``burst'' power by hand).  The total power in the short null average 
profile in Figure~\ref{nullprofiles} (comprised of the 296 remaining nulls) was 
nonzero---\ie, it did {\bf not} represent noise, rather there is very clear indication 
of emission.  This short null profile---showing clear emission features at the 
positions of the MP and IP---exhibits an aggregate intensity 0.18 $<$$I$$>$, a 
level well below the threshold but far above the noise fluctuations.

It is challenging yet of great importance to decipher whether this non-zero power 
is the result of emission from (a) consistent low intensity emission throughout the 
null sequences and/or (b) a small number of high-{\it er} intensity pulses that 
skew the average.  To reduce the possibility that the detected profile was due 
to (b), we reemphasize, the few pulses {(\bf half a dozen)} for which there was 
visible power (\ie, perhaps ``should'' have been called weak bursts) were removed 
by hand from the short null partial PS.  It was then after this procedure that 
significant power was detected in the short null profile (whereas, we will see below 
that the long null profile is indistinguishable from noise). We therefore conclude 
that the power in the short nulls is due to (a): consistent low intensity emission not 
detectable on a single pulse basis.

Short nulls are more or less randomly distributed throughout the PSs both at L 
and P band.  Recall that B1944+17's subpulses consistently repeat behaviors 
(evidenced by the star's well defined modes) though not with any perceptible 
regularity.  It is not surprising that the short nulls complimenting such modes appear 
consistently, but not periodically.

Mode changes are often, though not always, punctuated by short nulls.  Modes A 
and D are often interrupted by short nulls of some 1--3 $P_1$. Of the four modes, 
mode D is most frequently punctuated by short nulls. That the short nulls probably 
represent empty sightline traverses is very consistent with mode D's emission 
behaviors.  Mode C characteristically alternates short quasiperiodic bursts and 
nulls of some 10 -- 20 $P_1$. These transitions between nulls and bursts appear 
to entail no turnoff as no possible such ``partial null'' was ever observed in the 
pulse window.  Their characteristic length falls roughly on the boundary where we 
distinguish between short and long nulls. We have found no means of testing 
whether these mode-C nulls represent the longest pseudonulls or the shortest of 
the long null intervals.

\subsubsection{Long Nulls}
The rest of the story belongs to the long uninterrupted null sequences.  As mentioned 
earlier, B1944+17's nulls last up to 300 $P_1$ in our observations; however, note 
that we define long nulls to be any null sequences lasting longer than 7 rotation 
periods.  Long nulls appear very consistently in this pulsar, and their presence alone 
indicates an ``undermixing'' of bursts and nulls.  As discussed in \S 3 and displayed 
in Fig.~\ref{burstnullfreq}, the long null distribution is highly non-random.
 
In contrast to the short nulls, the aggregate power in the long nulls at P band 
is only 0.009 $<$$I$$>$, roughly that of noise fluctuations, demonstrating that 
there is truly no measurable power from the star during these nulls.  Every long 
null interval in our observations has a noise-like profile.  This is the case at both 
L and P band.

Considering their noise-like character, length, and non-random distribution, the 
probability that long nulls represent ``empty'' sightline traverses is extremely small.  
The evidence assembled indicates that long nulls represent actual cessations 
of the pulsar's emission.  This finding would be strengthened by measuring a 
decrease in spindown rate during the nulls, as in B1931+24; unfortunately, even 
the longest nulls in B1944+17 are far too short for this to be possible. 

In summary, we find that there are two distinct mechanisms behind B1944+17's 
nulls:  first, short {\it pseudo}nulls that are an artifact of carousel properties 
and rotation, exhibit measurable aggregate emission and tend to punctuate high 
intensity, well structured burst sequences; and second, long (true) nulls which 
are distributed non-randomly, have no detectable aggregate power, and thus 
plausibly represent actual cessations of the pulsar's emission engine.  In view of 
this star's varied emission behaviors, it is not surprising that its nulling phenomena 
are similarly complex.

\subsubsection{Null transitions}
Deich \etal\ looked for, but did not observe, transitions from the null-to-burst or 
burst-to-null state, that were expected to occur infrequently just as the star's 
emission cone swept through the Earth's direction---that is, occasions when the 
star's emission switched off or on {\it during} the pulse window.  In our investigation 
we found two such possible transitions at L band, one from the burst-to-null state 
and the other the opposite.

In this PS there were some 100 long null intervals, therefore some 200 transitions 
to and from long nulls.  As the MP profile has a roughly 30\degr\ equivalent width, 
we have at best a $\frac{1}{12}$ chance of observing a given transition.  This implies 
about 17 transitions; however we were able to identify only 2 with good confidence.  
This seeming shortage of null-burst transitions cannot be read into deeply:  a) it may 
be that such transitions can be securely identified only over a 10-15\degr\ interval, b) 
a number of the shorter long null intervals may not be what they seem, or c) confident 
detection may require particular modes before and after the long null.  And in any case, 
merely 2 partial nulls throws any speculation into the regime of small number statistics.

This all said, both transitions occurred approximately halfway through the pulse 
window and were marked by a sharp decrease or increase in emission intensity.  Both 
transitions occurred on the edges of long null intervals, from mode C in both cases.  
If these are valid indicators of transitions to or from cessations of emission, they occur 
on millisecond scales that are comparable to our longitude resolution.

\section{Summary and Discussion}
We have thoroughly investigated the radio pulse-sequence properties of pulsar 
B1944+17.  This star nulls some two-thirds of the time, has four distinct modes of emission 
and an IP, and we find that its magnetic and rotation axes are closely aligned.  

We confirmed the four modes first identified by Deich \etal.  Their properties are as 
follows:
\begin{itemize}
\item Mode A is the brightest and most ordered of the star's four modes.  It consistently 
displays three to four subpulses that drift with a $P_3$ value of about 14 periods. 
\item Mode B drifts at a faster rate and displays a $P_3$ of approximately 6 $P_1$.  
It usually shows three subpulses, which are characteristically less ordered than those 
of mode A. 
\item Mode C resembles mode A but is stationary.  It displays between one and three 
subpulses with similar brightness and virtually identical spacing ($P_2$) to those of 
mode A, however with negligible subpulse drift---and bursts alternate quasiperiodically 
with nulls of 10-30 $P_1$.  Mode C may be interpreted as representing a carousel of 
``beamlets'' organized in the same way as in mode A, but which is nearly stationary 
and flickering quasiperiodically. 
\item Mode D is the weakest of the four modes.  Though it lacks the ordered drift bands 
of the other modes, we find that it is not chaotic as it had been previously described by 
DCHR.  Its brightest subpulses are in the leading edge of the pulse window, appear 
consistently at the same phase, and remain active for some 5 periods or so. 
\item Both the total power and the FWHM as displayed in the modal average profiles 
show consistent behaviors {\it within} the different modes. 
\end{itemize}

In order to provide a sound basis for interpretation, we investigated B1944+17's 
emission geometry, and we drew the following conclusions:
\begin{itemize}
\item The pulsar has both an MP and IP, which occupy opposing roughly 60\degr\ 
intervals of its rotation cycle.  
\item The MP and IP are separated by almost exactly half a rotation period, independent 
of radio frequency.
\item The nearly complete linear polarization and disorderly PS modulation of the IP 
stands in stark contrast to the modest linear and rich phenomenology of the MP.  
\item The IP subpulses vary widely in intensity but show negligible correlation with 
subpulses in the MP.
\item The MP's equivalent width increases with increasing frequency, whereas its full 
width is nearly constant over a broad band.  This appears contrary to the frequency 
evolution exhibited by most other pulsars.
\item The PPA rates ($R$) of the star's MP and IP are unusually small, but of opposite 
senses.  The MP value is only +0.75\degr/\degr.  
\item Emission-beam modeling constrained by the MP and IP PPA-traverse information 
argues that $\alpha$ and $\beta$ are some 1.8 and 2.4\degr, respectively---such that the 
sightline orbit has an angular radius of just over 4\degr\ and remains inside the conal 
emission beam (See Fig.~\ref{beamgeometry}).
\item The small $\alpha$ and $\beta$ values indicate an unusual single-pole interpulse 
geometry---that is, here the IP emission occurs along field lines opposite to the MP and 
is produced by weak core emission. 
\item The rarity of B1944+17's geometrical configuration seems to account for some 
of its perplexing modulation and polarization phenomena.  
\end{itemize}

The prominent nulls of pulsar B1944+17 have been the principal interest driving this 
investigation.  Here, we summarize the patterns of its nulling behavior, discuss the 
null-length distribution, and evaluate which possible mechanisms are responsible for 
the nulls.
\begin{itemize}
\item B1944+17 appears to null nearly 70\% of the time.  That is, some two-thirds of the 
star's ``pulses'' have intensities falling below a plausible ``null threshold'' and are 
putative nulls.  In practical terms then, there is a two-thirds chance that any given ``pulse'' 
will be a ``null'' rather than a burst.
\item In the above context, it is remarkable to find that B1944+17 exhibits bursts 
lasting up to 100 or so $P_1$ and nulls with durations of up to some 300 periods.  
\item In Runs Test terms, the star's long nulls and bursts are non-randomly 
``undermixed''.  
\item One-period nulls appear with the highest frequency.  Nulls with lengths up to 
approximately 7 $P_1$ or so occur with a roughly random distribution; whereas 
nulls longer than this are obviously (and increasingly) non-random.
\item The short nulls---that is, those with durations less than roughly 8 $P_1$---have 
significant aggregate power in the form of a weak average MP and IP profile; see 
Fig.~\ref{nullprofiles}.  These short nulls are largely {\it pseudo}nulls---that is,  ``empty'' 
sightline traverses through the rotating carousel-beam system. 
\item Nulls longer than the above, on the other hand, exhibit negligible aggregate 
power and thus have the character of noise.  These long nulls then represent actual 
cessations of the pulsar's emission.  B1944+17's behavior is therefore similar to the 
intermittent pulsar B1931+24 (Kramer \etal\ (2006). 
\end{itemize}

On this basis, we can draw some interesting conclusions about the overall emission 
properties of pulsar B1944+17---
\begin{itemize}
\item Within certain A- or B-mode PSs, the emission from the pulsar's inner and 
outer cones exhibits similar modulation frequencies and thus folds synchronously.  
This strongly suggests that the inner and outer-cone emission is produced by the 
same set of particles.  
\item Similarly, the MP and IP emission generally bursts and nulls together, but is 
otherwise uncorrelated.  This suggests either that the core emission is produced 
at low altitude within the polar flux tube by sets of particles that produce the conal 
emission at higher altitude, or, perhaps that the conal emission is refracted inward 
as suggested by Petrova \& Lyubaski (2000).  
\item If we are correct about the star's emission geometry, then the star's subpulse 
modulation reflects a complex combination of inner and outer-cone contributions.  
\end{itemize}

Our analyses found no basis for determining the pulsar's carousel circulation time; 
however, some speculations are appropriate.  First, identification of a circulation 
time may be difficult or impossible in a pulsar with intermittent emission.  Second, 
rough estimations of the star's circulation time from Ruderman \& Sutherland (1975) 
or by extrapolaring from B0943+10 (DR01) both suggest that it could be short---\ie, 
10-15 $P_1$.  If there were then 10 or more ``beamlets'' in the carousel, the primary 
``drift'' modulation frequency could be aliased into the second order.  Such aliasing 
together with discrete changes in ``beamlet'' number could give rise to ``modes'' such 
as are observed in B1944+17.  For instance, were the star's carousel to circulate in 
some 12 $P_1$ with a configuration of 13 ``beamlets'', drift modulation similar to the 
A mode would be produced.  Similarly, 14 ``beamlets'' would modulate the PSs with 
a $P_3$ like that of the B mode,  and 12 would produce a nearly stationary modulation 
very like that of  mode C.  



We found it both fascinating and challenging to study a star such as B1944+17 
which exhibits so many distinct phenomena.  Clearly, its complex modes and nulls 
still have much to teach us.  Different polarization measurements may be able to 
better define its sightline geometry, and interferometry could well reveal a ``pedestal'' 
of continuous emission.  Finally, even longer observations are needed to further 
explore the star's carousel-beam configuration and fully assess the frequency 
of partial nulls.

\section*{Acknowledgments}
We are pleased to acknowledge Vishal Gajjar, Tim Hankins, Joeri van Leeuwen 
and Geoffrey Wright for their critical readings of the manuscript and Jeffrey Herfindal 
for assistance with aspects of the analysis.  We also sincerely thank both Tim 
Hankins for providing us with the original interpulse discovery observations, 
and Joel Weisberg for the ionospheric Faraday rotation corrections.  One of 
us (IMK) sincerely thanks the Barry M. Goldwater Scholarship and Excellence 
in Education program and the UVM College of Arts and Sciences for the APLE 
Summer Fellowship, which together permitted her to complete this work.  The other (JMR) 
thanks the Anton Pannekoek Astronomical Institute of the University of 
Amsterdam for their generous hospitality and the Netherlands National 
Science Foundation and ASTRON for her Visitor Grants.  Portions of this 
work were carried out with support from US National Science Foundation 
Grants AST 99-87654 and 08-07691.  Arecibo Observatory is operated by 
Cornell University under contract to the US NSF.  This work used the NASA 
ADS system.

{}

\end{document}